\magnification=1100
\baselineskip=15truept
\hsize=6.25truein
\vsize=8.75truein
\emergencystretch=2truepc\raggedbottom\frenchspacing\parskip=0pt
\font\Tfont=cmbx12
\def\T#1\par{\centerline{\Tfont#1}}
\newcount\secno\newcount\intro\newcount\causal\newcount\crsd
\newcount\piqm\newcount\mmm\newcount\syncor\newcount\clamac
\newcount\poties\newcount\corwcor\newcount\conclu
\newcount\ptff\newcount\wdmom\newcount\stdw
\def\insertsecno{\ifnum\secno=1I\fi\ifnum\secno=2II\fi
    \ifnum\secno=3III\fi\ifnum\secno=4IV\fi\ifnum\secno=5V\fi
    \ifnum\secno=6VI\fi\ifnum\secno=7VII\fi\ifnum\secno=8VIII\fi
    \ifnum\secno=9IX\fi\ifnum\secno=10X\fi\ifnum\secno=11XI\fi
    \ifnum\secno=12XII\fi\ifnum\secno=13XIII\fi\ifnum\secno=14XIV\fi
    \ifnum\secno=15XV\fi\ifnum\secno=16XVI\fi\ifnum\secno=17XVII\fi
    \ifnum\secno=18XVIII\fi\ifnum\secno=19XIX\fi\ifnum\secno=20XX\fi}
\def\secref#1{\ifnum#1=1I\fi\ifnum#1=2II\fi
    \ifnum#1=3III\fi\ifnum#1=4IV\fi\ifnum#1=5V\fi
    \ifnum#1=6VI\fi\ifnum#1=7VII\fi\ifnum#1=8VIII\fi
    \ifnum#1=9IX\fi\ifnum#1=10X\fi\ifnum#1=11XI\fi
    \ifnum#1=12XII\fi\ifnum#1=13XIII\fi\ifnum#1=14XIV\fi
    \ifnum#1=15XV\fi\ifnum#1=16XVI\fi\ifnum#1=17XVII\fi
    \ifnum#1=18XVIII\fi\ifnum#1=19XIX\fi\ifnum#1=20XX\fi}
\def\section#1#2\par{\advance\secno1#1=\secno\bigskip%
    {\bf\noindent\insertsecno.\ #2\par}\smallskip}
\def\longsection#1#2\par#3\par{\advance\secno1#1=\secno%
    \bigskip{\bf\noindent\insertsecno.\ #2\par
    \noindent#3\par}\smallskip}
\def\name#1\par{\vskip22truept\centerline{\rm#1}}
\def\affil#1\par{\vskip11truept\centerline{\it#1}}
\font\nnfont=cmr7\newcount\nnref\newcount\nnrefref
\def\nref{\advance\nnref1\raise1ex\hbox{\nnfont\number\nnref}}
\def\nrefref{\advance\nnref1\nnrefref=\nnref\advance\nnrefref1%
    \raise1ex\hbox{\nnfont\number\nnref,\number\nnrefref}%
    \advance\nnref1}
\def\oneref#1{\advance\nnref1\raise1ex\hbox{\nnfont\number#1,%
    \number\nnref}}
\def\nn#1{\nnrefref=\nnref\advance\nnref1\advance\nnrefref#1%
    \raise1ex\hbox{\nnfont\number\nnref-\number\nnrefref}%
    \advance\nnref#1\advance\nnref-1}
\def\onenn#1#2{\nnrefref=\nnref\advance\nnref1\advance\nnrefref#2%
    \raise1ex\hbox{\nnfont\number#1,\number\nnref-\number\nnrefref}%
    \advance\nnref#2\advance\nnref-1}
\def\repref#1{\raise1ex\hbox{\nnfont\number#1}}
\def\absosq#1{\bigl\vert#1\bigr\vert^2} 
\def\ket#1{\vert#1\rangle}

\def\sandwich#1#2#3{\langle#1\vert#2\vert#3\rangle}

\vglue1truein
\T The Pondicherry interpretation of quantum mechanics

\name Ulrich Mohrhoff\hskip2pt\raise1ex\hbox{\nnfont a)}

\affil Sri Aurobindo Ashram, Pondicherry 605002, India


\vskip30truept{\leftskip=\parindent\rightskip=\parindent\noindent
This article presents a novel interpretation of quantum mechanics. It extends 
the meaning of ``measurement'' to include all property-indicating facts. 
Intrinsically space is undifferentiated: there are no points on which a world of 
locally instantiated physical properties could be built. Instead, reality is built on 
facts, in the sense that the properties of things are extrinsic, or supervenient on 
property-indicating facts. The actual extent to which the world is spatially and 
temporally differentiated (that is, the extent to which spatiotemporal relations 
and distinctions are warranted by the facts) is necessarily limited. 
Notwithstanding that the state vector does nothing but assign probabilities, 
quantum mechanics affords a complete understanding of the actual world. If 
there is anything that is incomplete, it is the actual world, but its 
incompleteness exists only in relation to a conceptual framework that is more 
detailed than the actual world. Two deep-seated misconceptions are 
responsible for the interpretational difficulties associated with quantum 
mechanics: the notion that the spatial and temporal aspects of the world are 
adequately represented by sets with the cardinality of the real numbers, and 
the notion of an instantaneous state that evolves in time. The latter is an 
unwarranted (in fact, incoherent) projection of our apparent ``motion in time'' 
into the world of physics. Equally unwarranted, at bottom, is the use of causal 
concepts. There nevertheless exists a ``classical'' domain in which language 
suggestive of nomological necessity may be used. Quantum mechanics not 
only is strictly consistent with the existence of this domain but also 
presupposes it in several ways.\par}

\vfill\break
\section{\intro}INTRODUCTION

Following Mermin's recent example,\nref\ I propose to add another specimen to 
the quantum cabinet of curious interpretations. Mermin chose to call his 
specimen the Ithaca interpretation of quantum mechanics (IIQM). By the same 
naming scheme, what is presented in this article is the Pondicherry 
interpretation of quantum mechanics (PIQM).

Mermin tries to remove the mystery from quantum mechanics in just ten words: 
``Correlations have physical reality; that which they correlate, does not.'' He 
does not claim that there are no correlata, only that they are not part of {\it 
physical} reality. They belong to a larger reality which includes consciousness. 
According to the IIQM, the measurement problem arises in this larger reality, 
and it arises only when consciousness gets into the story. Being a puzzle about 
consciousness, it is not a proper subject for a physical theory.

I fully agree with Mermin that ``conscious perception... should be viewed as a 
mystery about {\it us} and should not be confused with the problem of 
understanding quantum mechanics.'' However, as I see it, the problem of 
understanding quantum mechanics either {\it is} the measurement problem or 
concerns the presuppositions that give rise to the measurement problem. It 
seems to me that in declaring the measurement problem to be a puzzle about 
consciousness, Mermin does precisely what he warns us against: he confuses 
the problem of consciousness with the problem of understanding quantum 
mechanics. The measurement problem ought to be solved, or shown to be a 
pseudoproblem, without dragging in conscious observers.

Quantum mechanics is, if nothing else, a tool for calculating probabilities. 
Mermin rightly insists that these probabilities are objective, in the sense that 
they have nothing to do with ignorance -- there is nothing for us to be ignorant 
of. I share his belief that all the mysteries of quantum mechanics can be 
reduced to the single puzzle posed by the existence of objective probabilities. 
What I remain skeptical about is his belief that one can nevertheless achieve a 
better understanding of quantum mechanics without squarely addressing this 
puzzle. To my mind, the problem of understanding quantum mechanics is as 
inseparable from the question of how statistical concepts like ``probability'' and 
``correlation'' can have ``meaningful application to individual systems,'' as it is 
from the measurement problem.

The article is organized as follows. Section~II shows that objective probabilities 
can be assigned only to counterfactuals. Objective probability distributions are 
distributions over the results of measurements that could have been, but were 
not, performed. Further it is shown that probabilities are objective if and only if 
they are calculated on the basis of all relevant facts, including those that are 
still in the future. This result is reinforced in Sec.~III, wherein it is argued that 
our apparent position and motion in time are as extraneous to physics as are 
our location and motion in space. Whatever is based on the intuitive notion of 
an advancing {\it now} (for instance, the distinction between the past and the 
future) has nothing to do with physics. Objective probabilities therefore must be 
time-symmetric. By the same token, backward-in-time causation must be as 
possible as forward-in-time causation. If the properties of material objects could 
be thought of as intrinsic, it would nevertheless be possible to interpret the 
physics without reference to backward causation. The objectivity of
quantum-mechanical probabilities, however, entails that the contingent 
properties of material objects are extrinsic rather than intrinsic, as is shown in 
Sec.~IV. This means that they are defined in terms of what happens or is the 
case in the rest of the world. Further it is shown in this section that, as a 
consequence, the multiplicity and the distinctions inherent in our mathematical 
concept of space cannot be intrinsic features of physical space. The notion that 
the spatiality of the world is adequately represented by a transfinite set of 
triplets of real numbers, is the principal fallacy preventing us from 
understanding how probability can be an objective feature of the world.

Section V addresses the widespread misconception that it is the business of 
quantum mechanics to account for the existence of facts. Quantum mechanics 
always presupposes, and never allows us to infer, the existence of facts. The 
state vector serves to assign probabilities to possibilities; it does not warrant 
inferences to actualities. Nor can it be thought of as an evolving collection of 
potentialities, for the time dependence of the state vector is a dependence on a 
stipulated time and not the time dependence of something that evolves in time. 
The very notion of an instantaneous state that evolves in time is an 
unwarranted (in fact, incoherent) importation from our successive experience 
into the world of physics, as is shown in Sec.~VI. Where the observed system 
is concerned, not only isn't there any actual state of affairs that obtains in the 
intermediate time between successive measurements, but also there isn't any 
intermediate time. The reason this is so is that times, like positions, are 
extrinsic: The actually existing times are the factually warranted times at which 
properties are possessed.

Section VII offers answers to a series of questions, posed by Mermin, 
concerning the role of measurements in quantum theory. The PIQM extends 
the meaning of ``measurement'' to include all property-indicating facts. These 
are external to the theory in the sense that the meanings of such locutions as 
``actual event'' and ``(matter of) fact'' cannot be defined in purely physical 
terms. This should not come as a surprise. The factuality of a fact is an 
ontologically concept as primitive as the spatiality of space or the temporality of 
time (or, for that matter, the subjectivity of consciousness). Section~VIII 
explains why quantum physics, but not classical physics, confronts us with this 
truism. While in classical physics actuality attaches itself to a nomologically 
possible world trivially through the initial conditions, in quantum physics it 
``pops up'' unpredictably and inexplicably with every property-defining fact. 
Section IX addresses the intriguing question of when a moving object moves. It 
is shown that the temporal referent of motion is a set of factually warranted 
moments that is not dense in time.

In Sec.~X macroscopic objects are rigorously defined -- as rigorously as is 
possible in view of the objective indefiniteness that is revealed by the existence 
of objective probabilities. Quantum mechanics not only is consistent with but 
also presupposes the existence of a classical domain. It is {\it quantitatively} 
consistent with quantum mechanics to think of the positions of macroscopic 
objects as forming a self-contained system of intrinsic properties that ``dangle'' 
causally from each other rather than ontologically from position-indicating facts. 
This is shown without positing ad hoc limits to the validity of quantum 
mechanics, and without compromising on the extrinsic nature of all contingent 
properties. Section~XI shows that, although no object ever possesses a sharp 
position, the ``fuzziness'' of the position of a macroscopic object exists only in 
relation to an imaginary background that is more differentiated spatially than is 
the actual world. That fuzziness therefore exists solely in our minds. This is 
what gives us the right to treat the positions of macroscopic objects as intrinsic, 
and to define all positions in terms of the positions of macroscopic objects. 
Considering that locations are not intrinsic to space (Sec.~IV), this is also the 
only way in which positions can be defined.

In Sec. XII two points are made. First, although at bottom causal concepts are 
nothing but anthropocentric projections deriving from our self-perception as 
agents, they work in the classical domain where the correlations between 
property-defining facts evince no statistical variations. (They had better work 
somewhere, for without a modicum of causality it wouldn't be possible to state 
the property-defining facts.) Second, quantum mechanics is inconsistent with 
local realism and the separability that this entails. There are no points on which 
a world of locally instantiated physical properties can be built. The world is built 
on facts, and its spatiotemporal properties are supervenient on the facts. To 
understand how EPR correlations are possible, one needs to understand that, 
in and of itself, physical space is undifferentiated. At a fundamental level, 
``here'' and ``there'' are the same place. If EPR correlations require a medium, 
this identity is the medium.

Section XIII concludes by contrasting the PIQM with the Copenhagen 
interpretation of quantum mechanics (CIQM). The CIQM stands accused of 
applying a double standard, treating measurements as physical processes 
governed by quantum mechanics, and again as constituents of a classical 
domain existing in an anterior logical relationship to quantum mechanics. By 
extending the meaning of ``measurement'' to include all property-indicating 
facts, the PIQM eliminates the double standard. The property-defining events 
are not governed by quantum mechanics; they are amenable to classical 
description, and this not merely ``for all practical purposes.'' A trait commonly 
attributed to the CIQM is an epistemic construal of the state vector: the latter 
represents our knowledge of the factual situation rather than the factual 
situation itself. The CIQM's claim to completeness is then understood as an 
agnosticism concerning the latter: quantum mechanics is complete in the sense 
that it enables us to say all that can be said with the language and the concepts 
at our disposal. According to the PIQM, quantum mechanics is complete in the 
sense that it enables us to say all that needs to be said in order to understand 
the actual world. If there is anything that is incomplete, it is the actual world. 
But its incompleteness exists only in relation to a conceptual framework that is 
more detailed than the actual world.

A glossary of technical terms is provided as an appendix.

\section{\corwcor}CORRELATIONS WITHOUT CORRELATA?

The case for the (physical) nonreality of the correlata is this: If one {\it 
assumes} that the correlations are real, the correlata cannot be real. If a 
composite system consisting of two spin-$1/2$ particles is (according to the 
usual phraseology) ``in the singlet state,'' no values can be assigned to the 
spin components of the individual particles. The existence of correlations, 
however, logically entails the existence of correlata. One cannot ascribe 
physical reality to the correlations and deny it to the correlata. If there is 
nothing that is correlated, there are no correlations either. This reduces to 
absurdity the assumption that the correlations (qua correlations between {\it 
objective} probability distributions) are physically real. If it cannot be the case 
that correlations and correlata are both physically real, then neither correlations 
nor correlata are physically real.

When both the correlations and the correlata are real in the literal (that is, {\it 
statistical}) sense of ``correlations'' and ``correlata,'' the correlata are actually 
possessed properties, and the correlations are correlations between statistical 
distributions over such properties. Statistical distributions are probability 
distributions only in the subjective sense of ``probability''; to assign probabilities 
to the possible results of actually performed measurements is to ignore the 
actual results. Objective probabilities cannot be assigned to actually possessed 
properties or to actually obtained measurement results. What sort of thing, 
then, is capable of being assigned an objective probability? What are objective 
probabilities distributed over? And what exactly are correlations that are not 
correlations between statistical distributions? The obvious answer to the first 
question is: a counterfactual. According to the PIQM, this is also the {\it only} 
correct answer. Nothing but a contrary-to-fact conditional can be assigned an 
objective probability. Objective probabilities are distributed over 
counterfactuals. Correlations between objective probability distributions are 
correlations between probability distributions over the possible results of {\it 
unperformed} measurements. Objective probabilities are objective in the sense 
that they are not subjective, and they are not subjective because they would be 
so only if the corresponding measurements were performed. In short, objective 
probabilities are probabilities that are {\it counterfactually subjective}.

This, then, is how we should understand the conclusion that neither the 
correlations nor the correlata are physically real: The correlata are properties 
that are not actually possessed, and the correlations are joint distributions over 
such properties. These distributions assign probabilities to counterfactuals (that 
is, to conditional statements that tell us nothing about what actually happens or 
is the case).

It is tempting to attribute the truth of objective probability assignments to an 
underlying actual state of affairs. It is equally tempting to assume that this 
actual state of affairs is somehow represented by the state vector. On this view, 
the singlet state
$\ket{0}\propto\ket{\uparrow\downarrow}-\ket{\downarrow\uparrow}$
not only serves to assign probabilities but also represents an actual state of 
affairs that accounts for the probabilities. However, if there were such an actual 
state of affairs, it would not be described by $\ket{0}$, for the singlet state does 
not describe anything; all it does is assign probabilities. The idea that a tool for 
assigning probabilities to {\it possibilities} describes an {\it actual} state of 
affairs is a category mistake. Moreover, if there were an underlying state of 
affairs, we would have to ask when and for how long it obtains, and it is well 
known that to this question there is no covariant 
answer.\nref\newcount\bohm\bohm=\nnref\ Finally, and most importantly, if 
there is a matter of fact about the value taken by a component of either of the 
two spins at any time after the ``preparation'' of the singlet state, the 
probabilities assigned by $\ket{0}$ are {\it not} objective, and the objective 
probabilities associated with the spin components are {\it not} given by 
$\ket{0}$.\nref

The reason this is so is that probabilities can be objective only if they are based 
on a complete set of facts. Otherwise they are subjective: They reflect our 
ignorance of some of the relevant facts. Born probabilities in general are 
calculated on the basis of an incomplete set of facts; they take into account the 
relevant past matters of fact but ignore the relevant future matters of fact. (By 
``Born probabilities'' I mean the probabilities associated with the standard 
formulation of standard quantum mechanics.) Born probabilities are objective 
only if there are no relevant future matters of fact.

ABL probabilities, on the other hand, also take into account the relevant future 
matters of fact. Named after Aharonov, Bergmann, and 
Lebowitz,\nref\newcount\abl\abl=\nnref\ ABL probabilities are calculated using 
a nonstandard formulation of standard quantum theory known as
time-symmetrized quantum theory.\nn{3}\newcount\av\av=\nnref\advance%
\av-2\ This time-symmetric formulation takes due account of the fact that the 
maximally specified state of a system contains information based not only on 
initial but also on final measurements. (If a measurement of the $x$ component 
of the spin of an electron, performed at $t_i$, yields $\uparrow_x$, and a 
measurement of the $y$ component of the spin of the same electron, performed 
at $t_f>t_i$, yields $\uparrow_y$, then a measurement of the $x$ component 
would with certainty have yielded $\uparrow_x$ if it had been performed at an 
intermediate time $t_m$, and a measurement of the $y$ component would with 
certainty have yielded $\uparrow_y$ if it had been performed at $t_m$.\nref)

Born probabilities can be measured (as relative frequencies) using preselected 
ensembles (that is, ensembles of identically ``prepared'' systems). ABL 
probabilities can be measured using pre- and postselected ensembles (that is, 
ensembles of systems that are both identically ``prepared'' and identically 
``retropared''). Both types of probability are assigned to possible results of 
possible measurements. If a possible measurement is actually performed, even 
the ABL probabilities are calculated on the basis of an incomplete set of facts; 
they take into account all revelant facts except the result of the actually 
performed measurement. ABL probabilities are based on a complete set of 
facts, and are therefore objective, only if none of the measurements to the 
possible results of which they are assigned is actually performed (that is, only if 
between the ``preparation'' or preselection and the ``retroparation'' or
postselection no measurement is performed).

Thus probabilities are objective only if they are distributed over alternative 
properties or values none of which are actually possessed, and only if they are 
based on all relevant matters of fact, including events that haven't yet 
happened and states of affairs that are yet to obtain.

\section{\causal}TIME AND CAUSALITY

If the objective probabilities associated with contrary-to-fact conditionals 
depend on events that haven't yet happened or states of affairs that are yet to 
obtain, some kind of retroactive causality appears to be at work. This 
necessitates a few remarks concerning causality and the apparent ``flow'' of 
time. But first let us note that nothing entails the existence of time-reversed 
causal connections between {\it actual} events and/or states of affairs. If at 
$t_m$ the $y$ component of the electron's spin is actually measured and the 
results at $t_i$ and $t_f$ are as specified above, nothing compels us to take 
the view that $\uparrow_y$ was found at $t_m$ {\it because} the same result 
was obtained at $t_f$. We can certainly stick to the idea that causes precede 
their effects, according to which $\uparrow_y$ was found at $t_f$ {\it because} 
the same result was obtained at $t_m$. The point, however, is that nothing in 
the physics prevents us from taking the opposite view. The distinction we make 
between causes and effects is based on the apparent motion of our location in 
time -- the present moment -- toward the future. This special location and its 
apparent motion are as extraneous to physics as are our location and motion in 
space. Equally extraneous, therefore, is the distinction between causes and 
effects.\nref\newcount\price\price=\nnref

Physics deals with correlations between actual events or states of affairs, 
classical physics with deterministic correlations, quantum physics with 
statistical ones. Classical physics allows us to explain the deterministic 
correlations (abstracted from what appear to be universal regularities) in terms 
of causal links between individual events. And for some reason to be explained 
presently, we identify the earlier of two diachronically correlated events as the 
cause and the later as the effect. The time symmetry of the classical laws of 
motion, however, makes it equally possible to take the opposite view that the 
later event is the cause and the earlier event the effect. In a deterministic world, 
the state of affairs at any time $t$ determines the state of affairs at any other 
time $t'$, irrespective of the temporal order of $t$ and $t'$. The belief in an 
exclusively future-directed {\it physical} causality is an animistic projection of 
the perspective of a conscious agent into the inanimate world, as I proceed to 
show.

I conceive of myself as a causal agent with a certain freedom of choice. But I 
cannot conceive of my choice as exerting a causal influence on anything that I 
knew, or could have known, at the time $t_c$ of my choice. I can conceive of 
my choice as causally determining only such events or states of affairs as are 
unknowable to me at $t_c$. On a simplistic account, what I knew or could have 
known at $t_c$ is everything that happened before $t_c$. And what is 
unknowable to me at $t_c$ is everything that will happen thereafter. This is the 
reason why we tend to believe that we can causally influence the future but not 
the past. And this constraint on {\it our} (real or imagined) causal efficacy is 
what we impose, without justification, both on the deterministic world of 
classical physics and on the indeterministic world of quantum physics.

When I decide on how exactly I should kick a football in order to score a goal, I 
use my knowledge of the time-symmetric law that governs the ball's 
trajectory.\nref\ I think of the kick as the cause and of the goal scored as its 
effect. This asymmetric causal relation has nothing to do with the
time-symmetric physics I exploit in order to produce the desired effect. It has 
everything to do with my self-perception as an agent and my successive 
experience of the world. My asymmetric agent causality rides piggyback on the 
symmetric determinisms of the physical world, and in general it rides into the 
future because in general the future is what is unknowable to me. But it may 
also ride into the past. Three factors account for this possibility.

First, as I said, the underlying physics is time-symmetric. If we ignore the 
strange case of the neutral kaon (which doesn't appear to be relevant to the 
interpretation of quantum mechanics), this is as true of quantum physics as it is 
of classical physics. If the standard formulation of quantum physics is 
asymmetric with respect to time, it is because we think (again without 
justification) that a measurement does more than yield a particular result. We 
tend to think that it also prepares a state of affairs which evolves toward the 
future. However, if this is a consistent way of thinking -- it is {\it 
not}\nref\newcount\umretro\umretro=\nnref -- then it is equally consistent to 
think that a measurement ``retropares'' a state of affairs that evolves toward the 
past, as transpires from the work of Aharonov, Bergmann, and 
Lebowitz.\repref{\abl}

Second, what matters is what can be known. If I could know the future, I could 
not conceive of it as causally dependent on my present choice. In fact, if I could 
(in principle) know both the past and the future, I could not conceive of myself 
as an agent. I can conceive of my choice as causally determining the future 
precisely because I cannot know the future. This has nothing to do with the 
truism that the future does not (yet) exist. Even if the future in some way 
``already'' exists, it can in part be determined by my present choice, provided I 
cannot know it at the time of my choice. By the same token, a past state of 
affairs can be determined by my present choice, provided I cannot know it 
before the choice is made.\nref

There are two possible reasons why a state of affairs $F$ cannot be known to 
me at a given time $t$: (i)~$F$ may obtain only after $t$; (ii)~at $t$ there may 
as yet exist no matter of fact from which $F$ can be inferred. This takes us to 
the last of the three factors which account for the possibility of retrocausation: 
{\it The contingent properties of physical systems are extrinsic}. By a {\it 
contingent} property of a system $S$ I mean a property that may or may not be 
possessed by $S$ at a given time. For example, being inside a given region of 
space and having a spin component of $+\hbar/2$ along a given axis are 
contingent properties of electrons. By an {\it extrinsic} property of $S$ I mean a 
property that is undefined, and hence nonattributable, unless its being 
possessed by $S$ can be inferred from what happens or is the case in the 
``rest of the world'' ${\cal W}-S$. A contingent property that is not extrinsic is 
intrinsic. A contingent property $p$ of $S$ is {\it intrinsic} if and only if the 
proposition ${\bf p}=$~``$S$~is~$p$'' is ``of itself'' (that is, unconditionally) 
either true or false at any time; neither the meaning of $\bf p$ nor the 
possession by $\bf p$ of a truth value depends on the goings-on in ${\cal W}-
S$.

Properties that can be retrocausally determined by the choice of an 
experimenter, cannot be intrinsic. If $p$ is an extrinsic property of $S$, the 
respective criteria for the truth and the falsity of $\bf p$ are to be sought in 
${\cal W}-S$, and it is possible that neither criterion is satisfied, in which case 
$\bf p$ is neither true nor false but meaningless. It is also possible that each 
criterion consists in an event that may occur only after the time to which $\bf p$ 
refers. If this event is to some extent determined by an experimenter's choice, 
retrocausation is at work. On the other hand, if $p$ is an intrinsic property of 
$S$, $\bf p$ possesses a truth value (that is, it is either true or false) 
independently of what happens in ${\cal W}-S$, so {\it a fortiori} it possesses a 
truth value independently of what happens there after the time $t$ to which $\bf 
p$ refers. There is then no fundamental reason why the truth value of $\bf p$ 
should be unknowable until some time $t'>t$. In principle it is knowable at $t$, 
and therefore we cannot (or at any rate, need not) conceive of it as being to 
some extent determined by the experimenter's choice at $t'$.

A paradigm case of retrocausation at work is the experiment of Englert, Scully, 
and Walther,\nrefref\ which I discussed in a previous article.\repref{\umretro}\ 
This experiment enables the experimenters to choose between (i)~measuring 
the phase relation with which a given atom emerges coherently from two slits 
and (ii)~determining the particular slit from which the atom emerges. They can 
exert this choice after the atom has emerged from the slit plate and even well 
after it has hit the screen. By choosing to create a matter of fact about the slit 
taken by the atom, they retroactively cause the atom to have passed through a 
particular slit. By choosing instead to create a matter of fact about the atom's 
phase relation, they retroactively cause the atom to have emerged with a 
definite phase relation. The retrocausal efficacy of their choice rests on the 
three factors listed above (in different order): (i)~The four propositions ${\bf 
a}_1=$~``the atom went through the first slit,'' ${\bf a}_2=$~``the atom went 
through the second slit,'' ${\bf a}_+=$~``the atom emerged from the slits in 
phase,'' and ${\bf a}_-=$~``the atom emerged from the slits out of phase'' affirm 
{\it extrinsic} properties. (ii)~There exist {\it time-symmetric} correlations 
between the atom's possible properties at the time of its passing the slit plate 
and the possible results of two incompatible (mutually exclusive) experiments 
that can be performed at a later time. (iii)~The result of the actually performed 
experiment is the first (earliest) matter of fact about either the particular slit 
taken by the atom or the phase relation with which the atom emerged from the 
slits. Before they made their choice, the experimenters could not possibly have 
{\it known} the slit from which, or the phase relation with which, the atom 
emerged.

Probabilities, I said, can be objective only if they are based on all relevant 
matters of fact, including events that haven't yet happened or states of affairs 
that are yet to obtain. We now see more clearly why it should be so. Our 
distinction between the past, the present, and the future, as Mermin likewise 
observes,\nref\ has nothing to do with physics. Physics ``knows nothing of {\it 
now},'' so it cannot know anything of the difference between the past and the 
future. An objective physical probability therefore cannot depend on a selection 
of facts that is based on this difference.

\section{\crsd}THE CONTINGENT REALITY OF SPATIAL DISTINCTIONS

The extrinsic nature of the contingent properties of physical systems is implied 
by the existence of objective probabilities. To see this, recall that objective 
probabilities are assigned to alternative properties none of which are actually 
possessed. Take the counterfactual ``If there were a matter of fact about the slit 
taken by the atom, the atom would have taken the first slit.'' We can assign to 
this counterfactual an objective probability (other than $0$ or $1$) only if the 
proposition ``The atom went through the first slit'' is neither true nor false but 
meaningless. But this can be the case only if the atom's whereabouts are 
extrinsic, for predications of intrinsic properties possess truth values at all 
times.

This result is a first significant step toward understanding probability as an 
objective feature of the physical world, rather than as a tactical device for 
coping with our ignorance. The existence of objective probabilities tells us that 
the contingent properties of things, in particular their positions, are extrinsic. 
And since ``[t]here is nothing in quantum theory making it applicable to three 
atoms and inapplicable to $10^{23}$,''\nref\ this must be as true of footballs and 
cats as it is of atoms and particles. The position of a material object $O$ is {\it 
defined} in terms of what happens or is the case in the rest of the world. It 
``dangles'' from actual events or states of affairs. $O$'s position is what matters 
of fact imply concerning $O$'s position.

The extrinsic nature of the contingent properties of physical systems has 
important implications. One of them is the contingent nature of spatial 
distinctions, as I proceed to show. Let $R$ be a region of space, and let $\bf a$ 
be the proposition ``$O$ is inside $R$.'' It is generally considered sufficient for 
the truth of $\bf a$ that the support of the retarded (``prepared'') wave function 
associated with $O$'s center of mass is contained in $R$. Considering the 
time-symmetry of the underlying physics, it ought to be equally sufficient for the 
truth of $\bf a$ that the support of the corresponding advanced (``retropared'') 
wave function is inside $R$. According to the PIQM, neither of these conditions 
is sufficient for the truth of $\bf a$. The necessary and sufficient condition for 
the presence of $O$ in $R$ at a time $t$ is the existence of a fact that indicates 
$O$'s presence in $R$ at $t$. If there isn't any such fact (at $t$ or at any time 
before or after $t$), and if there also isn't any event or state of affairs that 
indicates $O$'s absence from $R$ at $t$, then $\bf a$ is meaningless, and 
$O$'s position with respect to $R$ is objectively undefined -- where $O$ at the 
time $t$ is concerned, nothing in the world corresponds to the distinction 
between ``inside $R$'' and ``outside $R$.''

The conceptual distinctions we make between mutually disjoint regions of 
space thus may or may not be real for a given object at a given time. The 
distinction between $R$ and its complement $R'$ is real for $O$ if there is a 
matter of fact from which $O$'s position with respect to $R$ (inside $R$ or 
outside $R$) can be inferred; otherwise it is not. It follows that the multiplicity 
inherent in our mathematical concept of space (a transfinite set of triplets of 
real numbers) is {\it not} an intrinsic feature of the real thing, physical space. If 
the distinctions inherent in that concept were physically real {\it per se} (that is, 
if they were intrinsic to space), they would be real for every object that exists in 
space. In particular, the two slits in a double-slit experiment would be distinct 
for whatever passes through them, and so the individual atom could not pass 
through the (set-theoretic) union of the slits without passing through a particular 
slit {\it and} without being divided into distinct parts by its passage through the 
slits. But this is what it does when interference fringes are observed. The 
existence of the fringes proves that the slits are not distinct for the atoms, and 
that therefore spatial distinctions cannot be real {\it per se}.

According to the PIQM, the notion that the multiplicity and the distinctions 
inherent in our mathematical concept of space are intrinsic to physical space 
(and that, consequently, the individual points of space or space-time can be 
treated as the carriers of physical qualities) is a delusion. This notion perhaps 
more than any other prevents us from understanding how probability can be an 
objective feature of the physical world. It is as inconsistent with quantum 
mechanics as the notion of absolute simultaneity is with special relativity.

\longsection{\piqm}CAN QUANTUM MECHANICS ACCOUNT FOR FACTS?

CAN ANYTHING ELSE?

It is widely believed that it is the business of quantum mechanics to account for 
the occurrence/existence of actual events or states of affairs. From the point of 
view of the PIQM, this too is a misconception. Quantum mechanics assigns 
probabilities. These are determined by facts (the ``preparation'' and/or the 
``retroparation''), and they are assigned to possible facts. And the probability 
assignments are correct only if one assumes (in the case of objective 
probabilities, counterfactually) that one of a specified set of mutually exclusive 
possibilities is a fact.

To illustrate this point, consider two perfect detectors $D_1$ and $D_2$ having 
the respective (disjoint) sensitive regions $R_1$ and $R_2$. If the support of 
the wave function associated with the (center-of-mass) position of $O$ is 
neither wholly inside $R_1$ nor wholly inside $R_2$, nothing necessitates the 
detection of $O$ by $D_1$, and nothing necessitates the detection of $O$ by 
$D_2$. Yet if the wave function vanishes outside $R_1\cup R_2$, it is certain 
that either of the detectors will click. Two perfect detectors with sensitive 
regions $R_1$ and $R_2$ constitute one perfect detector $D$ with sensitive 
region $R_1\cup R_2$. But how can it be certain that one detector will click 
when individually neither detector is certain to click? What could cause $D$ to 
click while causing neither $D_1$ nor $D_2$ to click?

The answer is, nothing. That two perfect detectors with disjoint sensitive 
regions constitute one perfect detector for the union of their sensitive regions is 
part of what we {\it mean} by a perfect detector. By {\it definition}, a perfect 
detector clicks when the quantum-mechanical probability for it to click is~1. $D$ 
is certain to click because the respective probabilities for either detector to click 
add up to~1. Hence the question of what {\it causes} $D$ to click does not 
arise. If real detectors would behave like perfect detectors, it would be proper 
to inquire why they do so. But real detectors do not behave like perfect 
detectors. If the quantum-mechanical probability associated with finding $O$ in 
$R_1$ is $1/2$, a perfect detector with sensitive region $R_1$ clicks in 50\% of 
all runs of the experiment, while a real detector $D^{\rm\scriptstyle r}_1$ with 
the same sensitive region clicks in 50\% of all runs of the experiment {\it in 
which either detector clicks}. Hence the apparent commonplace that no (real) 
detector is perfect is not as trivial as it seems. Between a real detector and a 
perfect detector there exists not merely a quantitative difference that might one 
day be overcome; there exists a qualitative gap that no technological advance 
can bridge. The quantum-mechanical probability associated with finding $O$ in 
$R_i$, $i=1\hbox{ or }2$, is {\it conditional on} the existence of a matter of fact 
about the region containing $O$. Perfect detectors are so defined as to 
eliminate this conditionality; the definition stipulates that the condition is met. 
Hence what is a conditional probability for a real detector is an absolute 
probability for a perfect detector.\nref\newcount\detefcy\detefcy=\nnref\

Quantum mechanics thus tells us either of two things: (i)~{\it If} there is going to 
be a matter of fact about the alternative taken (from a specific range of 
alternatives) {\it then} such and such are the subjective probabilities with which 
that matter of fact will indicate this or that alternative. (ii)~{\it If} there were a 
matter of fact about the alternative taken (from a specific range of alternatives) 
{\it then} such and such are the objective probabilities with which that matter of 
fact would indicate this or that alternative. {\it Quantum mechanics always 
presupposes, and therefore never allows us to infer, the existence of a fact that 
indicates the alternative taken.} If the quantum-mechanical probability 
associated with finding $O$ in $R_1$ is~1, quantum mechanics accounts for 
the fact that whenever a detector clicks, it is $D^{\rm\scriptstyle r}_1$ that 
clicks. Quantum mechanics does {\it not} account for the fact that a detector 
clicks. And if quantum mechanics is as fundamental and as complete as I 
believe it is, there also isn't anything going on ``behind the back'' of quantum 
mechanics that accounts for the clicking.

Redhead\nref\newcount\redh\redh=\nnref\ has formulated the following 
``sufficiency'' condition: ``If we can predict with certainty, or at any rate with 
probability one, the result of measuring a physical quantity at time $t$, then at 
the time $t$ there exists an element of reality corresponding to the physical 
quantity and having a value equal to the predicted measurement result.'' This 
condition is {\it not} sufficient because quantum mechanics never tells us what 
will be the case, unconditionally. It only tells us what would be the case if some 
condition or other were met. Even if the Born probability of a particular event or 
state of affairs $F$ is~1, we are not entitled to infer that $F$ happens or 
obtains. Even the predictions of the standard formalism are {\it conditionals}. In 
order to get from a true conditional to an element of reality a condition has to 
be met: a measurement must be successfully performed, a matter of fact about 
the value of an observable must exist, one of a specified set of alternative 
property-defining events or states of affairs must happen or obtain.\nref

The failure to recognize and acknowledge the conditionality of
quantum-mechanical predictions has spawned an entire industry devoted to 
solving the so-called ``measurement problem.'' On the erroneous view that the 
quantum ``state'' represents an actual state of affairs, one is led to believe that 
there is a process of measurement which includes a ``premeasurement'' 
stage\nref\ during which the apparatus+system ``state''
$$
\ket{\Psi_1}=a\ket{N}\otimes\ket{a}+b\ket{N}\otimes\ket{b},\eqno{(1)}
$$
where $\ket{N}$ represents the neutral state of the apparatus, evolves into a 
state of the form
$$
\ket{\Psi_2}=a\ket{A}\otimes\ket{a}+b\ket{B}\otimes\ket{b}.\eqno{(2)}
$$
The measurement problem is the spurious problem of understanding the 
transition from the ``and'' of Eq. (1) or (2) to ``or'':
$$
\ket{A}\otimes\ket{a}\quad\hbox{or}\quad\ket{B}\otimes\ket{b}.\eqno{(3)}
$$
In truth, there is no such thing as a measurement {\it process} [that is, there is 
no actual state of affairs (1) that {\it evolves} into (2), there is no 
premeasurement {\it stage}, and there is no physical {\it transition} from (1) or 
(2) to (3)].\nref\ All that (1) tells us, is this: If there were a matter of fact from 
which the possession either of the property represented by $\ket{a}$ or of the 
property represented by $\ket{b}$ could be inferred, then that matter of fact 
would indicate the possession of $\ket{a}$ with probability $\absosq{a}$, and it 
would indicate the possession of $\ket{b}$ with probability $\absosq{b}$. And 
all that (2) tells us, is this: If there were two matters of fact, one from which the 
possession by the apparatus of either $\ket{A}$ or $\ket{B}$ could be inferred, 
and one from which the possession by the system of either $\ket{a}$ or 
$\ket{b}$ could be inferred, then the two matters of fact together would indicate 
the possession of both $\ket{A}$ and $\ket{a}$ with probability $\absosq{a}$, 
they would indicate the possession of both $\ket{B}$ and $\ket{b}$ with 
probability $\absosq{b}$, and they would indicate the possession of both 
$\ket{A}$ and $\ket{b}$ (or of both $\ket{B}$ and $\ket{a}$) with probability~0. It 
is self-evident that if there {\it is} a matter of fact from which either $\ket{a}$ or 
$\ket{b}$ can be inferred, and if this matter of fact is taken into account, the 
correct basis for further conditional inferences is either $\ket{a}$ or $\ket{b}$, 
depending on which of the two can be inferred. This obvious truism is the entire 
content of the projection postulate.\nrefref\newcount\vneu\vneu=\nnref\ There is 
no corresponding dynamical change of an actual physical state.

From the point of view of the PIQM, all attempts to coax ``classicality'' (actual 
events, matters of fact) out of quantum mechanics (via environment-induced 
superselection,\nn{5}\ decoherent histories,\nn{4}\ quantum state 
diffusion,\nn{3}\ spontaneous collapse,\nn{3}\ to mention but some) are 
misconceived and futile. Quantum mechanics takes us from facts to 
probabilities of possible facts; it cannot take us from possibilities to facts. The 
question of how it is that exactly one possibility is realized must not be asked of 
a formalism that serves to assign probabilities on the implicit {\it assumption} 
that exactly one of a specified set of possibilities {\it is} realized. Even the step 
from probability~1 to factuality crosses a gulf that quantum mechanics cannot 
bridge. The step from probability~1 (often wrongly paraphrased as ``certainty'') 
to a factually warranted inference takes us from a set of possible worlds (the 
framework in which counterfactuals are often discussed) to the real world. 
Quantum mechanics only takes us from the real world to the realm of possible 
worlds, and there it leaves us. But not only quantum mechanics fails to return 
us to reality. Out of that imaginary realm no logical road leads back to the real 
world. Measurements that are merely possible do not have actual results.

Left to its own resources, quantum mechanics appears to lead us into a 
Borgesian ``garden of forking paths.''\nref\ Every time a matter of fact about the 
state of a system $S$ comes into existence such as cannot be predicted with 
certainty on the basis of previous matters of fact, another object gets entangled 
with $S$, and the more objects are entangled with $S$, the more rapidly the 
entanglement spreads. The ``many-worlds extravanganza'' (to once again 
quote Mermin) gives shape to this fiction. What it overlooks is that the fiction is 
set in the realm of possible worlds. The purpose that the state vector serves in 
the {\it actual} world is to assign (Born) probabilities to conditional statements: 
{\it if} there is a matter of fact about the value of some observable, then such 
and such are the probabilities with which that matter of fact indicates the 
various possible values. But if there {\it is} any such matter of fact, those 
probabilities are based on an incomplete set of facts and are therefore 
subjective. All that ever gets {\it objectively} entangled is {\it counterfactuals}.

\longsection{\poties}DO MEASUREMENTS ACTUALIZE POTENTIALITIES?

THE MYTH OF AN EVOLVING INSTANTANEOUS STATE

The measurement problem is sometimes referred to as ``the problem of 
actualizing potentialities.''\nrefref\ The notion that quantum mechanics has 
something to do with the actualization of potentialities, which goes back to 
Heisenberg,\nref\ is misleading inasmuch as it suggests that the transition from 
potential to actual is a {\it process} of some kind (i.e., that it takes place {\it in 
time} and within a {\it single} world). If it is at all appropriate to think of a 
measurement in terms of a transition, this is not a transition in time (that is, not 
a physical transition from an earlier to a later state of affairs) but a ``transition'' 
from one possible world, in which a specified measurement is not performed at 
a stipulated time, to another possible world, in which the same measurement is 
performed at the stipulated time, as this section will show.

The idea that the state vector represents potentialities rather than an actual 
state of affairs is an improvement, but the idea that measurements actualize 
potentialities misses the point that {\it the time dependence of the state vector 
is a dependence on a stipulated time, not the dependence of something that 
evolves in time} -- not even if this is a collection of propensities\nref\ or a 
network of potentialities.\advance\nnref-4\nrefref\advance\nnref2\ Probabilities 
associated with events that may happen or states of affairs that may obtain, do 
not exist or evolve in time, anymore than the probability of being found inside a 
region $R$ is located inside $R$.

What is it that tricks us so persistently into thinking that there must be an 
instantaneous state that evolves in time? If there is such a state, then of course 
the quantum formalism leaves us no choice but to identify it with the
time-dependent state vector or density operator of the Schr\"odinger picture. 
But as soon as we do this, we are confronted with the pseudoproblem of why 
the state vector sometimes appears to unpredictably\nref\ ``collapse,'' and with 
the equally spurious problem of whether the ``collapse'' of the state vector is a 
real physical occurrence or something that happens only ``for all practical 
purposes.'' These problems are not solved by construing state-vector 
``collapses'' as transitions from possibilities to facts. They are ``solved'' by 
recognizing them as pseudoproblems, created by the fallacious notion of an 
instantaneous state that evolves in time, and the construal of the state vector 
as such a state.

To my mind, the root fallacy is the idea that the experiential now has anything 
to do with physics. The proper view of physical reality is not only what 
Nagel\nref\ has called ``the view from nowhere'' (physical reality is independent 
of the particular spatial location whence I survey the world); it is also what 
Price\repref{\price} has called ``the view from nowhen.'' Physical reality is 
independent of the particular time (the present) whence I survey the
spatiotemporal whole. As was shown in Sec.~\secref{\causal}, this entails that 
there is no place in physics for the qualitative difference that exists in 
conscious experience between the past, the present, and the future. It further 
entails that in this spatiotemporal whole there is no place for a ``moving'' now, a 
``flowing'' time, or an ``advancing'' present. In fact, the Whiteheadian 
concept\nref\ of an advancing present is incoherent, for the following reasons: it 
depicts space-time as a simultaneous whole persisting in a time that is 
extraneous to space-time, and it depicts the present as advancing through this 
persisting simultaneous whole in that extraneous time. In reality there is only 
one time, the fourth dimension of space-time. There is not another time in 
which the present ``advances'' through space-time as if space-time were a 
persisting simultaneous whole. Nor is it consistent to think of space-time as a 
simultaneous whole.

One may select any one-parameter family of spacelike hypersurfaces and call 
the parameter ``time'' and the hypersurfaces ``simultaneities.'' As far as physics 
is concerned, this association of states with times is all there is to time. One 
cannot pick a specific time and attach to the corresponding simultaneity the 
unique reality of the experiential now, let alone picture the now as advancing 
from simultaneity to simultaneity as if the simultaneities that are past or future 
relative to ``now'' existed simultaneously with the now. (These conclusions are 
independent of the fact that we live in a relativistic world. In a world in which 
simultaneities are a matter of choice, it is just more obvious that the unique 
reality of the experiential now cannot be attributed to any individual 
simultaneity.\nref)

If the concept of an advancing present is an unwarranted (in fact, incoherent) 
importation from our successive experience into the world of physics, then so is 
the concept of an advancing instantaneous state. This concept is part of a 
package of ``folk'' conceptions about time and causality that we must discard if 
we want to understand what quantum mechanics is trying to tell us. Here is how 
the ``folk tale'' goes: Since the future is not yet real, it cannot influence the 
present, so retrocausation is impossible. And since the past is no longer real, it 
can influence the present only by persisting through time (that is, by the 
persistence right up to the present of something that was real, however it may 
have changed in the meanwhile). Causal influences reach from the nonexistent 
past into the nonexistent future by being ``carried through time'' by the present. 
There is an evolving instantaneous state, and this not only represents presently 
possessed properties but also encapsulates everything in the past that is 
causally relevant to the future. This is how we come to conceive of ``fields of 
force'' that evolve in time (and therefore, in a relativistic world, evolve according 
to the principle of local action), and that ``mediate'' between the past and the 
future (and therefore, in a relativistic world, between local causes and their 
distant effects). It is also how we come to believe that the state vector plays a 
similar causally mediating role. It is high time that we outgrow these 
unwarranted (in fact, incoherent) beliefs.

If we find that at times $t_1$ and $t_2$ the system has the properties 
represented by $\ket{\psi_1(t_1)}$ and $\ket{\psi_2(t_2)}$, respectively, we 
learn nothing about the properties possessed by it at other times. We learn 
nothing about an evolving actual state of affairs that obtains in the meantime. If 
there isn't any matter of fact about what $S$ is like in the meantime, then there 
isn't anything that $S$ is like in the meantime. Hence where $S$ is concerned, 
there isn't any actual state of affairs that obtains in the meantime. There is no 
need for filling the temporal gap between $t_1$ and $t_2$ with an evolving 
instantaneous state, nor are we justified in doing so.

It stands to reason that what is true of the multiplicity inherent in our 
mathematical concept of space, is also true of the multiplicity inherent in our 
mathematical concept of time: neither multiplicity is real {\it per se}. Neither is 
space a storehouse of preexistent positions, nor is time a storehouse of 
preexistent instants. The actually existing positions are not the positions of 
``the points of space'' but the factually warranted positions of material objects. 
By the same token, the actually existing times are not ``the moments of time'' 
but the times of actual events or states of affairs or the factually warranted 
times at which properties are possessed. It follows that if there isn't any 
intermediate time at which $S$ has a factually warranted property, or if (where 
$S$ is concerned) no actual state of affairs obtains at any intermediate time, 
then (where $S$ is concerned) there just isn't any intermediate time. In 
particular, there isn't any time at which one could attribute to $S$ a collection of 
propensities or a network of potentialities.

The transition from a superposition of the form $\sum_i a_i\ket{a_i}$ to one of 
the kets $\ket{a_i}$, therefore, cannot be a transition from a potential state of 
affairs that obtains ``just before'' the measurement to an actual state of affairs 
that obtains at the time of the measurement (and on to another potential state 
of affairs that obtains ``just after'' the measurement). Where the system is 
concerned, the times ``just before'' and ``just after'' the measurement do not 
exist. The difference between $\sum_i a_i\ket{a_i}$ and a particular ket 
$\ket{a_i}$ is not a difference in time and within the same world but a difference 
between possible worlds at the same time. In a world in which $\sum_i 
a_i\ket{a_i}$ is the correct basis for (prior) probability assignments to 
counterfactuals, the measurement of the observable whose eigenstates are the 
kets $\ket{a_i}$ is not performed at the stipulated time. In those worlds in which 
it is performed at the stipulated time, the property represented by one of the 
kets $\ket{a_i}$ is actually possessed at that time, and the corresponding ket is 
the correct (partial) inference basis for probability assignments to 
counterfactuals pertaining to earlier or later times.

Consider once more a system $S$ that is found to possess the properties 
represented by $\ket{\psi_1(t_1)}$ and $\ket{\psi_2(t_2)}$ at the respective 
times $t_1$ and $t_2$. If there isn't any matter of fact about what $S$ is like in 
the meantime, we can say that $S$ has {\it changed} from an object that has 
the properties represented by $\ket{\psi_1(t_1)}$ into an object that has the 
properties represented by $\ket{\psi_2(t_2)}$ -- but {\it only} in the sense that at 
$t_1$ the system has the former properties and at $t_2$ the system has the 
latter properties. There is no interpolating state of affairs that evolves from a 
state in which $S$ has the former properties into a state in which $S$ has the 
latter properties.\nref\ The change {\it consists} in the fact that the properties of 
$S$ at $t_2$ differ from the properties of $S$ at $t_1$. Nothing can be said 
about the meantime, not because $S$ is propertyless in the meantime, but 
because, where $S$ is concerned, {\it there isn't any meantime}. Reality is not 
built on a space and a time that are infinitely and intrinsically differentiated; 
reality is built on matters of fact, and the actually existing positions and times 
are the factually warranted positions of material objects and the factually 
warranted times at which properties are possessed by such objects. ``When'' 
there is no factually warranted property, there is no factually warranted time, 
and ``when'' there is no factually warranted time, there is no time -- at least 
where $S$ is concerned.

\section{\mmm}THE MEANING OF ``MEASUREMENT''

Mermin asks: (1)~``Why should the scope of physics be restricted to the 
artificial contrivances we are forced to resort to in our efforts to probe the 
world?'' (2)~``Why should a fundamental theory have to take its meaning from a 
notion of `measurement' external to the theory itself? Should not the meaning 
of `measurement' emerge from the theory, rather than the other way round?'' 
(3)~``Should not physics be able to make statements about the unmeasured, 
unprepared world?'' The answers to these questions are as follows.

(1) If by ``measurement'' we mean a manipulation that is {\it intended} to 
determine the value of a given observable or that leads to the acquisition of {\it 
knowledge}, the scope of physics is not restricted to measurements. What is 
relevant is the occurrence or existence of an event or state of affairs warranting 
the assertability of a statement of the form ``$S$ has the property $p$ at the 
time $t$,'' irrespective of whether anyone is around to assert, or take 
cognizance, of that event or state of affairs, and irrespective of whether it has 
been anyone's intention to learn something about $S$. Bohr insisted that 
quantum systems should not be thought of as possessing properties 
independently of experimental arrangements.\nref\ By interpreting his 
sagacious insistence on the necessity of describing quantum phenomena in 
terms of experimental arrangements\nrefref\newcount\atdn\atdn=\nnref\ as 
restricting quantum mechanics ``to be exclusively about piddling laboratory 
operations,'' one does him an injustice. For ``experimental arrangement'' read: 
matters of fact about the properties possessed by the system at a given time. A 
``measurement result,'' properly understood, does not have to be the outcome 
of a laboratory experiment. {\it Any} matter of fact that ``is about'' (has a bearing 
on) the properties of a physical system, qualifies as a measurement result.

(2) The (proper) notion of ``measurement'' is external to the theory in the sense 
that locutions such as ``actual event,'' ``actual state of affairs,'' ``matter of fact'' 
cannot be defined in quantum-mechanical terms. This view is not likely to be 
popular with theoretical physicists who naturally prefer to define their concepts 
in terms of the mathematical formalism they use. Einstein spent the last thirty 
years of his life trying (in vain) to get rid of field sources, those bits of ``stuff'' 
that have the insolence to be real by themselves rather than by courtesy of 
some equation.\nref\ Small wonder if he resisted Bohr's insight that not even 
the {\it properties} of things can be defined in purely mathematical terms. But 
that's the way it is. The properties of things are objectively defined in terms of 
what actually happens or obtains, and the meaning of ``what actually happens 
or obtains'' is beyond the reach of quantum-mechanical definition. This point is 
important enough to be made in a section of its own -- the next.

(3) Physics is able to make statements about the unmeasured world, but {\it 
only in the terms of the measured world} (that is, only in terms of 
counterfactuals) -- perhaps with one exception. While in the first place quantum 
mechanics is about statistical correlations among facts, to a certain extent it 
seems to let us infer an underlying reality. Suppose that we perform a series of 
position measurements, and that every position measurement yields exactly 
one result (that is, each time exactly one detector clicks). Then we are entitled 
to infer the existence of an entity $O$ which persists through time (if not for all 
time), to think of the clicks given off by the detectors as matters of fact about 
the successive positions of this entity, to think of the behavior of the detectors 
as position measurements, and to think of the detectors as detectors. The lack 
of transtemporal identity among particles of the same type of course forbids us 
to extend to such particles the individuality of a fully ``classical'' entity. If each 
time exactly two detectors click, and if no distinguishing properties are 
detected, the following is neither true nor false but meaningless: A particular 
click at one time and a particular click at another time indicate the presence of 
the {\it same} particle.

On closer examination we find that the ``underlying'' reality -- in our examples a 
single entity $O$ or a system of exactly two particles -- ``dangles'' as much 
from the facts as do the positions that we attribute to its ``constituents''. There 
is a determinate number of objects {\it because} every time the same number of 
detectors click, and if we take the possibility of pair creation/annihilation into 
account, there is a determinate number of objects only {\it at} the times at which 
they are detected. In other words, not only the properties of things but also the 
number of existing things supervenes on the facts. The true constituents of 
reality therefore are not things but facts.

\longsection{\ptff}PHYSICAL THEORY AND THE FACTUALITY OF FACTS:

A SECOND LOOK

Quantum mechanics never predicts that a measurement will take place, nor 
does it predict the time at which one will take place, nor does it specify the 
conditions in which one will take place. The PIQM takes quantum mechanics to 
be complete in the sense that these things simply cannot be predicted or 
specified: Nothing {\it necessitates} the existence of a property-defining fact. In 
other words, a matter of fact about the value of an observable is a causal 
primary. (A {\it causal primary} is an event or state of affairs the occurrence or 
existence of which is not necessitated by any cause, antecedent or otherwise.)

I do not mean to say that in general nothing causes a measurement to yield this 
particular value rather than that. Unless one postulates hidden variables, this is 
a triviality. What I mean to say is that nothing ever causes a measurement to 
take place. Measurements (and in clear this means detection events) are 
causal primaries.\nref\ Quantum physics is concerned with correlations 
between events or states of affairs that are uncaused and therefore 
fundamentally inexplicable. As physicists we are not likely to take kindly to this 
conclusion, which may account for the blind spot by which its inevitability has 
been hidden so long. To the best of my knowledge, Mermin is the first who has 
bitten the bullet, for by denying physical reality to the correlata, he in effect 
declares that, at least where physics is concerned, the existence of the 
correlata (that is, the existence of factually warranted properties or of
property-defining facts) {\it is} fundamentally inexplicable.

Mermin thinks that there nevertheless must be something that accounts for the 
actuality of actual events and states of affairs, and, like several other quantum 
theorists,\onenn{\vneu}{5}\newcount\albert\albert=\nnref\newcount\lockw%
\lockw=\nnref\advance\lockw-1\ he believes that consciousness has something 
to do with it. As I see it, the idea that the factuality of facts needs to be 
accounted for has its roots in an inappropriate way of thinking about 
possibilities. Here are the possibilities to which quantum mechanics assigns 
probabilities, so where is the source of the actuality that some of them are 
blessed with? The fallacy lies in the antecedent. Where is ``here,'' and what 
does ``are'' mean? We must not let the copula ``are'' fool us into thinking that 
possibilities enjoy an existence of their own. It would be unnecessary to point 
out such an obvious category mistake, were it not for the widespread habit of 
endowing the ``state'' vector with a reality of its own. If one thinks of a tool for 
assigning probabilities to possibilities as a state of affairs that evolves in time, 
one needs something that is ``more actual'' than this state of affairs -- 
something capable of bestowing ``a higher degree of actuality'' -- in order to 
explain why every successful measurement has exactly one result. If one needs 
something ``more actual'' than the state vector, then consciousness is an 
obvious candidate. But it seems to me more straightforward to acknowledge 
that there is only one kind of actuality, that possibilities are just possibilities, 
and that the factuality of facts is not something that needs to be accounted for, 
any more than we are required to give an answer to the question of why there 
is anything at all, rather than nothing.

A comparison with classical physics might help to make this point. In classical 
physics one has, on one side, a theoretical description or account of the world 
and, on the other side, the world. The latter possesses something that every 
theory, concept, or description lacks -- it's {\it real}. Yet nobody expects a 
classical theory to account for the realness of the world. Nobody expects a 
classical theory to define this peculiar ``property'' of the world. The very idea of 
theoretically defining or accounting for the realness of the real is inherently 
absurd. How could one explain why there is anything rather than nothing? How 
could any theory account for or define the factuality of facts? This is an 
ontological concept as primitive as the spatiality of space, the temporality of 
time, the subjectivity of consciousness, or the blueness of blue. (Actually it's 
more primitive, for if there aren't any facts, then nothing is here, now, 
conscious, or blue.) And this is true independently of whether or not the world 
is classical.

The essential difference between classical and quantum physics lies in their 
respective ways of relating theory, possible worlds, and the one actual world. 
Classical physics {\it describes} (nomologically) possible worlds, and it 
describes them as causally closed. The name of the game is to find the laws 
according to which any given set of initial conditions, if it were realized, would 
evolve deterministically. If a particular set of initial conditions is realized, so is 
the entire future evolving from it in accordance with the classical dynamical 
laws. If a particular set of initial conditions is not realized, neither is the future 
determined by these initial conditions. The determinism of classical physics 
ensures that a possible classical world is either actual {\it en block} or not 
actual at all. The relation between possible worlds and the actual world is 
therefore trivial: The actual world is just one of the possible worlds described by 
classical physics.

Quantum physics describes neither (nomologically) possible worlds nor the 
actual world. We do not have, on one side, theoretical descriptions of possible 
worlds and, on the other side, the actual world. What we have is, on one side, 
rules for assigning objective probabilities to counterfactuals and, on the other 
side, a totality of uncaused facts, on the basis of which objective probabilities 
are assigned to counterfactuals. (If future facts are ignored, those rules can 
also be used to assign subjective probabilities to the results of measurements 
that are yet to be made.) Thus in quantum physics facts and counterfactuals 
are inextricably entwined. Quantum theory takes us from facts to probabilities 
of possibilities, and this has created the impression that it is incomplete. There 
ought to be something that takes us back to the facts. But this is an error. 
Actuality comes in statistically correlated but causally disconnected ``bits and 
pieces'' that just happen to be. Nothing can ``take us to'' a causal primary. At 
bottom the factuality of the facts is as ineffable as the reality that attaches itself 
to one of the possible worlds of classical physics, but now it looms large 
because it no longer enters through initial conditions (which, except in 
cosmology, are of no particular interest) but instead ``pops up'' unpredictably 
and inexplicably with every property-defining fact.

\section{\wdmom}WHEN DOES A MOVING OBJECT MOVE?

If property-defining facts are uncaused, we cannot say that they are {\it 
created}. Hence, strictly speaking, it wasn't correct to say, as I did in 
Sec.~\secref{\causal}, that the experimenters can choose between creating a 
matter of fact about the slit taken by the atom and creating a matter of fact 
about the atom's phase relation. All they can do is create conditions that permit 
the existence of either matter of fact, and optimize the likelihood of its existence 
by optimizing the efficiency of the corresponding apparatus.\nref\ 
Measurements are not {\it made}; they are only made possible or likely. By the 
same token, particles do not {\it trigger} detectors (that is, they do not {\it 
cause} them to respond); at most they enable them to respond, with a likelihood 
that depends on both the ``preparation'' of the particles and the efficiency of the 
detectors. Conversely, detectors do not {\it localize} particles (that is, they do 
not {\it cause} them to be in a specific region of space); at most they enable 
them to be in a specific region of space, with a likelihood that again depends on 
both the ``preparation'' of the particles and the efficiency of the detectors. Thus 
causal terms are out of place not only where the interfactual relations are 
concerned, but also where the relations between the decisions of 
experimenters, the property-defining facts, and the properties of things are 
concerned.

Where the properties of things are concerned, even temporal language is 
inappropriate. Physical systems do not {\it acquire} properties when a
property-defining event occurs. They do not {\it change} in response to a 
measurement. Nothing {\it happens} to $S$ at $t_2$ if $S$ is found to possess 
at $t_2$ the properties represented by $\ket{\psi_2(t_2)}$. The measurement 
warrants the system's having these properties at $t_2$, and that's all. Nothing 
justifies the idea that $S$ acquires these properties at $t_2$ or that its 
properties change at $t_2$. The only legitimate change that can be attributed 
to $S$ consists in the fact that its properties at $t_2$ differ from its properties at 
the preceding actual moment $t_1$, as we saw in Sec.~\secref{\poties}. This 
change takes place neither at $t_1$ nor at $t_2$, nor does it take place in the 
(for $S$) nonexistent interval between $t_1$ and $t_2$.

So when does a quantum system change? In particular, when does it move 
(change its position)? This question is reminiscent of Zeno's famous third 
paradox.\nref\ Zeno's argument goes like this: A flying arrow is in a particular 
location at every instant; at no instant does it change its location; therefore it is 
altogether at rest. Modern calculus enables us to resolve the paradox by 
treating motion as a local property -- but only in a classical world. The ``local'' 
property is defined by a limiting process that involves vanishing intervals of 
space and time, and in the actual world such a limiting process has no physical 
meaning. Nature's answer to Zeno has to be different. Zeno's conclusion is 
right: If time were a ``continuous'' set of instants (that is, if a finite time span 
contained an infinite set of moments at which the arrow has factually warranted 
positions), the arrow could not move, as the quantum effect named after 
Zeno\nn{3}\ demonstrates. What is wrong is the premise. Where the arrow is 
concerned, time is a discrete set of actual moments at which the arrow 
possesses factually warranted positions. The finite time span between any two 
consecutive such moments is nonexistent for the arrow. What allows the arrow 
to move is the gaps in the arrow's ``experience'' of time -- the gaps between 
moments that are real for the arrow. This doesn't mean that the arrow moves {\it 
during} those gaps. To move is to be in different places at different times. The 
temporal referent of the arrow's motion ``between'' two consecutive actual 
moments is not the (nonexistent) interval between these moments but the pair 
of moments itself. The temporal referent of the change that consists in the 
properties of $S$ at $t_2$ being different from the properties of $S$ at $t_1$ is 
the pair of moments $\{t_1,t_2\}$. This is also the temporal referent of the 
acquisition, by $S$, of the properties represented by $\ket{\psi_2(t_2)}$.

\section{\clamac}MACROSCOPIC OBJECTS

While the Moon isn't only there when somebody looks,\nref\ it is there only 
because of the myriads of matters of fact about its whereabouts.
Position-indicating matters of fact are constitutive of the Moon's being there. 
This seems to entail a vicious regress. We infer the positions of particles from 
the positions of the detectors that click. But the positions of detectors are 
extrinsic, too. They are what they are because of the matters of fact from which 
one can (in principle) infer what they are. This means that there are detector 
detectors from which the positions of particle detectors are inferred, and then 
there are detectors from which the positions of detector detectors are inferred, 
and so on {\it ad infinitum}. Again, the properties of things ``dangle'' 
ontologically from what happens or is the case in the rest of the world. Yet what 
happens or is the case there can only be described by describing material 
objects, and their properties too ``dangle'' from the goings-on in the rest of the 
world. This seems to send us chasing the ultimate property-defining facts in 
never-ending circles. Somewhere the buck must stop if the PIQM is to be a 
viable interpretation of quantum mechanics. In this section I show that the buck 
does stop, without positing ad hoc limits to the validity of quantum mechanics, 
and despite the fact that all contingent properties are extrinsic.

The existence of objective position probabilities tells us that the positions of 
things are objectively indefinite or ``fuzzy.'' This does not mean that $O$ {\it 
has} a fuzzy position. It means something like this: one can always conceive of 
a sufficiently small region $R$ of space such that the proposition ``$O$ is in 
$R$ at $t$'' not only lacks a truth value but also lacks a trivial probability (a 
probability of either 0 or~1). However, there are objects, which I will call 
``macroscopic,'' the positions of which are not {\it manifestly} indefinite.

Let me explain. A {\it macroscopic object} $M$ is an object that satisfies the 
following criterion: every factually warranted inference to the position of $M$ at 
any given time $t$ is predictable on the basis of factually warranted inferences 
to (i)~the positions of $M$ at earlier times and (ii)~the positions of other objects 
at $t$ or earlier times. (By saying that a factually warranted inference to the 
position of a macroscopic object is predictable, I do not mean that the existence 
of the position-indicating fact is predictable, but that the position indicated by 
the fact is predictable.) To put it negatively, nothing ever indicates that $M$ has 
a position different from what is predictable on the basis of the pertinent 
classical laws and earlier position-defining facts. Thus we may say that the 
positions of macroscopic objects are not {\it manifestly} indefinite, in the sense 
that the indefiniteness in their positions is not evidenced by the existence of 
unpredictable facts. Every matter of fact about $M$'s present position follows 
via the pertinent classical laws from matters of fact concerning $M$'s past 
position and the past and present positions of other objects.

The notion that the unpredictability of a position-indicating fact reveals a 
positional indefiniteness, requires some care. The position of an object $O$ 
can be indefinite with respect to a region $R$ only if the proposition $\bf 
a=$~``$O$ is in $R$'' has a nontrivial probability assigned to it. In this case 
there exist two possibilities: (i)~If there is a matter of fact about the truth value 
of $\bf a$, this is unpredictable, but its unpredictability does not reveal $O$'s 
positional indefiniteness with respect to $R$. If $\bf a$ has a truth value, the 
probability associated with $\bf a$ is subjective, so $O$'s position is not 
indefinite with respect to $R$. (ii)~If there isn't any matter of fact about the truth 
value of $\bf a$, $O$'s position is indefinite with respect to $R$, but this 
positional indefiniteness is not revealed by any unpredictable matter of fact. 
What {\it is} revealed by the unpredictability of an actual position-indicating 
event $e$ is a {\it counterfactual} indefiniteness: the indefiniteness that would 
have obtained had $e$ not occurred (other things being equal).

Note that the definition of a ``macroscopic object'' does not stipulate that events 
indicating random departures\nref\ from the classically predicted positions 
occur with zero {\it probability}. An object is entitled to the label ``macroscopic'' 
if no such event {\it actually} occurs. What matters is not whether such an event 
{\it may} occur (with whatever probability) but whether it ever {\it does} occur.

Now, an unpredictable matter of fact about the position of $O$ can exist only if 
there are detectors whose sensitive regions are small and localized enough to 
probe the space over which the position of $O$ is distributed. The existence of 
such a matter of fact entails the existence of detectors $D_i$, $i=1,\dots,n$, 
with sufficiently ``sharp'' positions. (A detector -- more precisely, an $O$-
detector -- is anything that is capable of warranting the presence of $O$ in a 
particular region of space.) The existence of an unpredictable matter of fact 
about the position of any of the detectors $D_i$ in turn entails the existence of 
detectors $D_{ik}$, $k=1,\dots,m$, whose sensitive regions are small and 
localized enough to probe the space over which the position of $D_i$ is 
distributed, and so on. It stands to reason that one sooner or later runs out of 
detectors with sharper positions. There will be ``ultimate'' detectors the 
positions of which are too sharp to be manifestly fuzzy.

Let me say this again. If $O$'s position is to be manifestly indefinite, there must 
exist detectors capable of probing the space over which the position of $O$ is 
distributed. To probe this space, they must have positions that are sharper than 
$O$'s position, and sensitive regions that are smaller than the space over 
which $O$'s position is distributed. But detectors with sharper positions and 
sufficiently small sensitive regions do not always exist. There is a finite limit to 
the sharpness of the positions of material objects, and there is a finite limit to 
the spatial resolution of real detectors. There are objects whose positions are 
the sharpest in existence. Hence not every object can have a manifestly 
indefinite position. Macroscopic objects exist. We cannot be certain that a 
given object qualifies as macroscopic, inasmuch as not all matters of fact about 
its whereabouts are accessible to us. But we can be certain that macroscopic 
objects exist, and that the most likely reason why $M$ is macroscopic is the 
nonexistence of detectors with positions that are sharper than the position of 
$M$, and with sensitive regions that are smaller than the space over which 
$M$'s position is distributed. ($M$ could also be macroscopic for the unlikely 
reason that such detectors, though they exist, never indicate a departure from 
$M$'s classically predicted position.)

Another word of caution. We must desist from thinking of $O$'s position as 
being distributed (``smeared out'') in any actual sense. Saying that $O$'s 
position is distributed over a set of mutually disjoint regions $\{R_i\vert 
i=1,\dots,n\}$, is the same as saying that the probability of finding $O$ in $R_i$ 
is positive for several or all values of $i$. If the probabilities are subjective, 
what is distributed (in the literal statistical sense) is an ensemble of identically 
``prepared'' systems. If the probabilities are objective, $O$'s position is not 
actually but {\it counterfactually} distributed (that is, it is distributed over 
contrary-to-fact conditionals). But if $O$'s position is counterfactually 
distributed over a set of mutually disjoint regions, there isn't any matter of fact 
about the particular region that contains $O$. In this case the multiplicity and 
the distinctness of those regions are not real for $O$ (Sec.~\secref{\crsd}), so 
there is nothing over which $O$ is {\it actually} distributed.

We are now in a position to resolve the apparent vicious regress pointed out at 
the beginning of this section. The buck does stop. There are ultimate detectors; 
there are objects whose positions are not manifestly indefinite because they 
are the ``most localized'' objects in existence. Although no object ever follows a 
definite trajectory, there are objects whose positions evolve in a completely 
predictable fashion. This makes it possible to think of the positions of such 
objects as forming a self-contained system of positions that ``dangle'' causally 
from each other, rather than ontologically from position-indicating facts. We can 
treat the positions of macroscopic objects as if they were intrinsic, without ever 
risking being contradicted by the implications of an actual event or state of 
affairs. The Moon is there because of the myriad of facts that betoken its 
presence, but its position ``dangles'' from them in a way that is predictable, a 
way that does not reveal any (counterfactual) fuzziness. It is therefore 
completely superfluous to apply to the positions of macroscopic objects the 
language of statistics, to assign to them probabilities, and to treat them as 
extrinsic.

Matters of fact about the positions of macroscopic objects are correlated in a 
way that permits us to project our asymmetric agent causality into the
time-symmetric world of physics (Sec.~\secref{\causal}), and to think of the 
positions possessed at later times as causally determined by the positions 
possessed at earlier times. This isn't the way it really is. In reality there are no 
causal links. It nevertheless is not {\it quantitatively} wrong to think of the 
positions of macroscopic objects as forming a self-contained system of intrinsic 
properties tied to each other by causal strings, and to quantitatively define all 
positions in terms of the not manifestly indefinite positions of macroscopic 
objects. There aren't any sharper positions in terms of which positions could be 
defined.

Bohr insisted not only on the necessity of describing quantum phenomena in 
terms of the experimental arrangements in which they are displayed, but also 
on the necessity of employing classical language in describing these 
experimental arrangements.\oneref{\atdn}\ The existence of a ``classical 
domain'' -- a domain adequately described in classical language -- is perfectly 
consistent with quantum mechanics. What Bohr was stressing is that the actual 
events or states of affairs presupposed by quantum mechanics cannot be 
described without referring to the (changeable) properties of persistent objects. 
(Try to describe an experiment without talking about persistent objects and their 
changing properties!) Thus, on the one hand, quantum mechanics requires the 
use of classical language -- the language of persistent objects and properties 
that to some extent evolve predictably. (Unless properties evolve in a 
sufficiently predictable fashion, they cannot be thought of as the properties of a 
persistent object because they cannot be thought of as the properties of the {\it 
same} object at different times.) On the other hand, the use of this language is 
consistent with quantum mechanics, despite the extrinsic nature of all 
contingent properties. Macroscopic objects exist, and it is legitimate to describe 
them in the language of persistent objects and causally connected properties. 
The use of causal terms is legitimate not because there is anything that {\it 
forces} macroscopic objects to evolve in a certain manner, but because the 
factually warranted positions of macroscopic objects are predictable -- they do 
not evince any indeterminism.

Note that an apparatus pointer is not a macroscopic object according to the 
above definition. In general nothing allows us to predict which way the pointer 
will deflect (given that it will deflect). However, {\it before} and {\it after} the 
deflection the pointer behaves as a macroscopic object, and this is sufficient for 
its initial and final positions to be defined independently of what happens 
elsewhere. Hence the deflection event -- the unpredictable transition from the 
initial to the final position -- is also defined independently of the rest of the 
world. The system property indicated by the deflection event explicitly 
presupposes the deflection event, but the latter is sufficiently described in 
terms of intrinsic properties; in order to describe it, we need not refer to other 
property-defining events.

The deflection event is part of the history of an object that is adequately 
described in classical language. Such an object can be thought of as 
possessing a persistent reality of its own, and events in its history can be 
thought of as deriving their actuality from this persistent reality. Hence there is 
a viable alternative to thinking of the actuality of the deflection event as 
``popping up'' unpredictably and inexplicably. The event happens unpredictably 
and inexplicably, but it owes its reality (as distinct from its particular nature) to 
the persistent reality of the pointer. While it is fundamentally correct that 
actuality -- the actuality of possessed properties -- comes in ``bits and pieces'' 
that are statistically correlated but not causally linked, the possessed properties 
of macroscopic objects, and even of pointer needles, are so correlated that 
their actuality can be grounded in the continuous reality of a persistent object.

\longsection{\stdw}THE SPATIAL DIFFERENTIATION OF THE WORLD:

FACT AND FICTION

How real is the indefiniteness of a position that is not manifestly indefinite? The 
answer is, {\it not real at all}. The ``most objective'' way for the position of an 
individual object $O$ to be distributed over a set of mutually disjoint regions 
$\{R_i\vert i=1,\dots,n\}$, recall, is to be {\it counterfactually} distributed. If 
$O$'s position is so distributed, there isn't any matter of fact concerning the 
particular region $R_i$ that contains $O$. The reason why there isn't any such 
matter of fact is either the absence of suitable detectors or their failure to 
indicate the position of $O$. If $O$ is not a macroscopic object, either such 
detectors could have been in place but were not, or they were in place but none 
did respond. In either case the multiplicity and the distinctness of the regions 
$\{R_i\}$ are not real for $O$. On the other hand, if $O$ is macroscopic, the 
most likely reason why there isn't any matter of fact about the particular region 
containing $O$ is that there are no detectors capable of probing the space over 
which $O$'s position is distributed. In this case the multiplicity and the 
distinctness of the regions $\{R_i\}$ fail to be real not only for $O$ but for every 
object in existence. If there are no detectors to realize the distinctness of those 
regions, then nothing in the actual world corresponds to our conceptual 
distinction between those regions.

We have this inveterate tendency to visualize everything against an infinitely 
and intrinsically differentiated backdrop. Since spatial distinctions are not 
intrinsic to space (Sec.~\secref{\crsd}), such a backdrop does not exist. The 
actually existing spatial distinctions are those that are warranted by facts. Like 
all contingent properties, they ``dangle'' from what happens or is the case. And 
since there is a finite limit to the spatial resolution of real detectors 
(Sec.~\secref{\clamac}), a finite region $R$ of space contains at most a finite 
number of regions that are actually distinct. For any object $O$ located within 
$R$, at most a finite number of distinct positions (``inside $R_i$,'' where 
$\{R_i\}$ is a partition of $R$) are available as possible attributes.

Suppose that $\{R_i\}$ is a partition of $R$ at the limit of resolution achieved by 
actually existing detectors. How should we visualize this partition? Certainly not 
as a set of sharply bounded regions! Since no object ever possesses an exact 
position, no detector ever has a sharply bounded sensitive region. Even the 
boundaries of the ultimate detectors are fuzzy. But since they cannot be 
manifestly fuzzy,\nref\ the mental picture of fuzzily bounded regions is more 
detailed than the finitely differentiated reality it is supposed to represent. The 
fuzziness exists only in relation to a more differentiated backdrop, and this 
exists only in our imagination. Where the regions $\{R_i\}$ or the sensitive 
regions of ultimate detectors are concerned, nothing in the actual world 
corresponds to the spatial distinctions implicit in the notion of a fuzzy boundary 
-- a boundary distributed (``smeared out'') over a multiplicity of mutually disjoint 
regions. By the same token, nothing in the actual world corresponds to the 
spatial distinctions implicit in the indefiniteness of the position of a macroscopic 
object. And this is the same as saying that nothing in the actual world 
corresponds to the indefiniteness of the position of a macroscopic object. This 
indefiniteness exists solely in our minds.

It seems to me that we must reconcile ourselves to the existence of objects the 
positions of which are neither sharp nor fuzzy, and which evolve neither in a 
deterministic, causal fashion nor in an unpredictable, indeterministic fashion. 
The notion that ``sharp'' and ``fuzzy'' are jointly exhaustive terms has its roots 
in an inadequate theoretical representation of the world's actual spatial 
determinations. That notion rests on the assumption of an intrinsically and 
infinitely differentiated space, and this assumption is inconsistent with quantum 
mechanics (Sec.~\secref{\crsd}). In a world that is spatially differentiated only to 
the extent that spatial relations and distinctions can be inferred from the facts, 
no object has a sharp position. But there are objects that have the sharpest 
positions in existence, and these positions are not fuzzy in any real sense; they 
are fuzzy only in relation to an unrealized degree of spatial differentiation. Their 
fuzziness exists exclusively in our heads.

If space is neither intrinsically nor infinitely differentiated, there are no points 
that could serve as the substrate for physical qualities. There are no points on 
which a world of locally instantiated properties could be built. This is why reality 
is built on facts (Sec.~\secref{\poties}), and why the spatial properties of things 
(including the actual spatial differentiation of the world) are extrinsic -- they 
``dangle'' from the facts. The extrinsic nature of possessed properties makes 
room for objective probabilities (Sec.~\secref{\crsd}), and these evince 
themselves counterfactually, through the unpredictability of some factually 
warranted properties. Quantum mechanics is our tool for calculating those 
probabilities. And if quantum mechanics is complete, what is unpredictable 
(and therefore uncaused) is not just the outcomes of some successfully 
performed measurements but the existence of every property-defining fact 
(Sec.~\secref{\ptff}). This implies that nothing behaves the way it does out of 
nomological necessity (compelled by causal laws) (Secs.~\secref{\poties}\ and 
\secref{\wdmom}). There are nevertheless things that evolve predictably, in the 
sense that their factually warranted positions are predictable on the basis of 
earlier position-defining facts. Thus there are objects that evolve neither 
deterministically (by nomological necessity) nor unpredictably.

\longsection{\syncor}EPR CORRELATIONS AND THE MYSTERY OF ACTION

AT A DISTANCE

Perhaps the most impenetrable aspect of the physical world consists in the 
correlations that are observed on systems in spacelike separation. The 
paradigm example of such correlations is (Bohm's\repref{\bohm}\ version of) 
the experiment of Einstein, Podolsky, and Rosen 
(EPR).\nref\newcount\epr\epr=\nnref\ Suppose that at the time $t_0$ two spin-
$1/2$ particles are in the singlet state (that is, there is a matter of fact by which 
this inference is warranted). At $t_1>t_0$ Alice measures the spin component 
of the first particle with respect to some axis, and at $t_2>t_0$ Bob measures 
the spin component of the second particle with respect to the same or a 
different axis. The temporal order of $t_1$ and $t_2$ is irrelevant. Whenever 
Alice and Bob choose the same axis, their results are perfectly anticorrelated.

Can there be a causal explanation for this phenomenon? That the
common-cause explanation fails is well known. Nor is agent causality involved, 
for two reasons. First, the correlations are perfectly symmetric. If Alice and Bob 
obtain the respective results $\uparrow^A_x$ and $\downarrow^B_x$, the 
following statements are equally true: (i)~Alice obtained $\uparrow^A_x$ {\it 
because} Bob obtained $\downarrow^B_x$ (and because she measured the 
$x$ component), (ii)~Bob obtained $\downarrow^B_x$ {\it because} Alice 
obtained $\uparrow^A_x$ (and because he measured the $x$ component). At 
the same time both statements are overspecific. If there is no objective criterion 
to decide which of two correlated events is the cause and which is the effect, 
causal concepts are inappropriate. Second, Bob's choice exerts no causally 
determining influence on the result obtained by Alice (and vice versa): no 
matter which axes they choose, the odds that Alice will obtain $\uparrow$ or 
$\downarrow$ are always fifty-fifty. Einstein was absolutely right: ``the real 
factual situation of the system $S_2$ is independent of what is done with the 
system $S_1$.''\nref

Once again, the concepts of action and causation are out of place. Diachronic 
correlations that are not manifestly indeterministic can be passed off as causal 
explanations. We can impose on them our agent causality with some measure 
of consistency, even though this compels us to use a wrong criterion: temporal 
precedence takes the place of causal independence as the criterion which 
distinguishes the cause from the effect. But when we deal with EPR 
correlations or correlations that are manifestly indeterministic, the imposition of 
agent causality does not work. Trying to causally explain these correlations is 
putting the cart in front of the horse. It is the correlations that explain why 
causal explanations work to the extent they do. They work in the classical 
domain where we are dealing with macroscopic objects, and where the 
correlations between property-indicating facts evince no statistical variations 
(dispersion). In this domain we are free to use language suggestive of 
nomological necessity. But if we go beyond this domain, we realize that all 
regularities are essentially statistical, even when statistical variations are not in 
evidence, and that our belief in nomological necessity is an unwarranted 
anthropocentric projection deriving from our self-perception as agents.\nref\ 
What lies beyond the classical domain is out of bounds to the concept of 
causation. The correlations are fundamental. There is no domain of underlying 
causal processes. There is a domain in which causal terms can be used, but 
this is defined by the specific condition that the statistical correlations evince no 
variations. It is the correlations that account for the limited usefulness of causal 
language. There is no underlying causation that could explain the correlations.

And still one wants to know how it can be like that\nref\ -- not how EPR 
correlations {\it work} but simply how they are {\it possible} at all. To answer 
this question, one must point out the fallacy that makes them seem impossible. 
As I see it, the fundamental stumbling block is the view known as {\it local 
realism} and the separability that this entails. According to local realism, ``...all 
there is to the world is a vast mosaic of local matters of particular fact, just one 
little thing and then another.... We have geometry: a system of external 
relations of spatiotemporal distance between points.... And at those points we 
have local qualities: perfectly natural intrinsic properties which need nothing 
bigger than a point at which to be instantiated.... And that is all.... All else 
supervenes on that.''\nref\ If an atom can pass through a double slit without 
passing through either slit in particular and without being divided into parts by 
its passage through the slits, the slits cannot demarcate spatial locations that 
are intrinsically distinct (Sec.~\secref{\crsd}). Neither space nor time are 
intrinsically differentiated. There are no points on which a world of locally 
instantiated properties can be built. The world is built on facts 
(Sec.~\secref{\poties}), and its spatiotemporal properties (including the actual 
extent of its spatial and temporal differentiation) are supervenient on the facts. 
In and of itself, physical space -- or the reality underlying it -- is 
undifferentiated.

Newton's law of gravity created a problem concerning which Newton himself 
refused to ``frame hypotheses'': How is action at a distance possible? Field 
theories are often hailed as providing the solution to the perceived paradox of 
unmediated action across space. But what enables field theories to do so is 
their commitment to the principle of local action (a.k.a. local causality), 
according to which the ``local matters of particular fact'' at a point $\cal P$ are 
influenced only by the ``local matters of particular fact'' in the infinitesimal 
neighborhood of $\cal P$ -- as if the electromagnetic interaction were explained 
by, rather than contributed to explain, what goes on when a hammer strikes a 
nail. Since space-time points are not intrinsically distinct, the principle of local 
action is without application, and the field theoretic ``solution'' is no solution. 
Nor is a solution needed, for the problem of action at a distance arises only if 
space-time points are conceived as intrinsically distinct. The same applies to 
the problem of how EPR correlations are possible. If in and of itself ``here'' is 
not distinct from ``there,'' there is no problem. What happens or is the case 
``here'' can be correlated with what happens or is the case ``there'' because, 
fundamentally, ``here'' and ``there'' are identical, not in the qualitative sense of 
exact similarity, but in the strict sense of numerical identity. At bottom there is 
only one place, and it is everywhere. If action at a distance or EPR correlations 
require a medium, that identity (or this ubiquity) is the medium.

\section{\conclu}CONCLUSION: PONDICHERRY VERSUS COPENHAGEN

Here is a list of wishes that many theoretical physicists harbor and, 
simultaneously or alternatingly, know to be chimerical: ``Suppose that there's 
just one world. And suppose that there's just one complete story of the world 
that's true. And suppose that quantum-mechanical state vectors are complete 
descriptions of physical systems. And suppose that the dynamical equations of 
motion are always exactly right.''\nref\ The PIQM endorses the first two items 
and rejects the remaining two. State vectors are not complete descriptions of 
physical systems because they do not {\it describe} physical systems 
(Sec.~\secref{\ptff}). Nor can one say that the dynamical equations (in the 
simplest case, the Schr\"odinger equation) are always (and everywhere) 
exactly right, given the usual acceptations of ``always'' and ``everywhere'' (``at 
every instant of time'' and ``at every point of space,'' where time and space are 
conceived as sets with the cardinality of the real numbers). Once again, the 
spatial properties of things and the times at which they are possessed exist to 
the extent that they can be inferred from facts, and facts do not warrant the 
inference of an infinitely differentiated space or time. Between the factually 
warranted times that make up the history of a system, there are gaps; there are 
``periods of time'' that do not exist for the system (Sec.~\secref{\poties}). 
Everything that can be said about those periods is counterfactual. The state 
vector assigns (prior) probabilities to the possible results of measurements that 
are {\it not} performed during such a period, or else it assigns subjective 
probabilities to the possible results of the measurement that is performed at the 
end of the period. The time dependence of the state vector is a dependence 
either on the time of an unperformed measurement, which time is nonexistent 
for the system, or on the time of the measurement that is performed at the end. 
The ``dynamical'' equations are ``always'' exactly right only in the sense that 
they correctly spell out this time dependence.

The CIQM\nref\ equally denies that state vectors are descriptions of physical 
reality, and that they always evolve in accordance with the ``dynamical'' 
equations. It further agrees with the PIQM in that the state vector is a tool for 
calculating the probabilities of measurement results, that values can be 
attributed only to observables that are actually measured, and that 
``measurement'' must be treated as a primitive concept, incapable of physical 
analysis. The PIQM goes beyond the CIQM in that it extends the meaning of 
``measurement'' to include all property-indicating facts. For the PIQM, the 
primitive concept is ``(matter of) fact.'' This eliminates the CIQM's Janus-faced 
portrayal of measurements: as physical processes governed by quantum 
mechanics, and again as constituents of a classical domain existing in an 
anterior logical relationship to quantum mechanics. The property-indicating 
events or states of affairs are not governed by quantum mechanics. They are 
amenable to classical description, in terms of objects whose positions behave 
exactly like deterministically evolving intrinsic positions (in the sense that 
nothing indicates that they don't) -- except when they themselves indicate the 
possession, by another object, of a property that wasn't predicted by the 
classical dynamics on the basis of the relevant earlier facts. Even pointer 
positions are extrinsic; they too would not be attributable if they couldn't be 
inferred from the (classically describable) goings-on in the rest of the world. 
Even the properties of the classical domain ``dangle'' from a myriad facts in the 
classical domain, but in a way that allows us to ignore the network of inferences 
from facts to properties which supports the classical domain, and to imagine in 
its stead a network of causal connections between intrinsic properties.

Strictly speaking, there is no such thing as {\it the} CIQM. Often but not always 
the CIQM is associated with a subjective or epistemic construal of 
``measurement'': The state vector represents our knowledge of the factual 
situation rather than the factual situation itself.\nref\ The CIQM's claim to 
completeness is then construed as an agnosticism concerning the factual 
situation: what is responsible for the experimental results and their statistical 
correlations is forever beyond our ken. Here is how Stapp\nref\ has 
characterized the conceptual innovation due to the CIQM: ``The theoretical 
structure did not extend down and anchor itself on fundamental microscopic 
space-time realities. Instead it turned back and anchored itself in the concrete 
sense realities that form the basis of social life.'' The PIQM is in full agreement 
with the first part of this statement, but it rejects the pragmatism of the second 
part. Science is driven by the desire to know how things {\it really} are. It owes 
its immense success in large measure to the belief that this can be discovered. 
There is no need to trade this powerful ``sustaining myth''\nref\ for the 
pragmatic notion that physics deals only with perceived and communicable 
phenomena. The theoretical structure is anchored in facts.

According to Stapp, ``[t]he rejection of classical theory in favor of quantum 
theory represents, in essence, the rejection of the idea that external reality 
resides in, or inheres in, a space-time continuum. It signalizes the recognition 
that `space,' like color, lies in the mind of the beholder.'' The PIQM fully agrees 
with the first part of this statement but rejects the subjectivism of the second 
part. Surely there are other ways of thinking about the spatial and temporal 
aspects of the world. The fact that reality is not built on an intrinsically and 
infinitely differentiated space does not imply that there is no objective space. 
Spatiality and temporality are objective aspects of the world, but the world is 
spatially and temporally differentiated only to the extent that spatial and 
temporal distinctions are warranted by the facts.

Stapp's {\it non sequitur} is symptomatic of one of the two deep-seated 
misconceptions that prevent us from understanding quantum mechanics -- the 
idea that the set-theoretic description of space with its inherent multiplicity and 
distinctions provides the {\it only} possible way of thinking about the spatiality 
of the world. (The other misconception is the notion of an evolving 
instantaneous state.) If this were true, the external world would be nonspatial 
and hence definitely beyond our ken. Quantum mechanics would be complete 
in the sense that it enables us to say all that can be said with the language and 
the concepts at our disposal. According to the PIQM, quantum mechanics is 
complete in the sense that it enables us to say all that needs to be said in order 
to understand the material world. (This is very different from the notion that the 
state vector itself describes the material world.) What is incomplete is not 
quantum mechanics but reality itself; that is, reality is incomplete relative to a 
description of reality that is ``overcomplete.'' The seemingly intractable problem 
of understanding quantum mechanics is a consequence of our dogged 
insistence on obtruding onto the world, not a spatiotemporal framework, but a
spatiotemporal framework that is more detailed than the world.

We think of the statistical regularities quantum mechanics is concerned with as 
statistical {\it laws} -- as being what they are {\it of necessity}. (Otherwise we 
could not employ quantum-mechanical probability assignments 
counterfactually; we could not assign {\it objective} probabilities to unperformed 
measurements.) But this necessity cannot be the nomological necessity of a 
causal law. So what sort of necessity is it? Is there something ``behind'' the 
statistical regularities, something that is responsible for or revealed by them? 
According to the PIQM, the answer is positive: what is revealed is an objective 
indefiniteness in the positions of nonmacroscopic objects. It has been said that, 
although the terminology of indefinite or fuzzy values is prevalent in some 
elementary textbooks, what is really intended is that a certain observable does 
not possess a value at all.\nref\ But there is more than that to the indefiniteness 
of a position. The statistical laws warrant assigning objective probabilities; they 
allow us to associate with an individual object an objective probability 
distribution over mutually disjoint regions of space. Hence they warrant the 
notion of an objectively indefinite position. Indefiniteness is an irreducible 
feature of the world. Qualitatively this is easy enough to grasp, although giving 
it exact quantitative expression is tricky. It involves (apart from renunciation of 
two deep-seated misconceptions about space and time) objective probabilities, 
counterfactuals, the extrinsic nature of contingent properties -- in fact, the 
entire conceptual apparatus put together in this paper.

\bigskip\bigskip\noindent{\bf APPENDIX: GLOSSARY}

\smallskip
{\it ABL probability}. The probability\repref{\av}
$P_{ABL}(a_i)={\absosq{
\sandwich{\Psi_2}{{\bf P}_{A=a_i}}{\Psi_1}
}\over
\Sigma_j\absosq{
\sandwich{\Psi_2}{{\bf P}_{A=a_j}}{\Psi_1}
}}$
with which a measurement of the observable $A$ performed on a system $S$ 
between the ``preparation'' of the state $\ket{\Psi_1}$ and the ``retroparation'' 
of the state $\ket{\Psi_2}$ yields the result $a_i$. The operator ${\bf 
P}_{A=a_i}$ projects on the subspace corresponding to the eigenvalue $a_i$ of 
$A$. ABL probabilities are based on past and future matters of fact about the 
properties of $S$.

{\it Agent causality}. The time-asymmetric causality of the goal-directed 
activities of a conscious agent in a successively experienced world. It is often 
wrongly superimposed on the time-symmetric causal links of classical physics 
and the time-symmetric relation in quantum physics between measurements 
and probability assignments.

{\it Born probability}. The probability $P_B(a_i)=|\sandwich{\Psi}{{\bf 
P}_{A=a_i}}{\Psi}|$ with which a measurement of the observable $A$ on a 
system $S$ performed after the so-called ``preparation'' of the state 
$\ket{\Psi}$ yields the result $a_i$. The operator ${\bf P}_{A=a_i}$ projects on 
the subspace corresponding to the eigenvalue $a_i$ of $A$. Born probabilities 
are based only on past matters of fact about the properties of $S$.

{\it Causal primary}. An event or state of affairs the occurrence or existence of 
which is not necessitated by any cause, antecedent or otherwise.

{\it Classical domain}. A collective name for the positions of macroscopic 
objects. 

{\it Classical language}. Quantum mechanics presupposes not only the 
existence of facts but also the possibility of describing facts in the classical 
language of persistent objects and causally connected properties. Despite the 
extrinsic nature of all contingent properties, the use of classical language is 
consistent with quantum mechanics. Macroscopic objects are adequately 
described in this language -- if one keeps in mind that the causal terminology is 
warranted by predictability rather than by nomological necessity.

{\it Conditional}. A hypothetical statement; a compound statement of the form 
``If $p$ then $q$.'' Component $p$ is called the antecedent.

{\it Contingent property}. A property that a system $S$ may, but does not 
necessarily, possess. Being inside a given region of space and having a spin 
component of $+\hbar/2$ along a given axis are contingent properties of 
electrons.

{\it Contrary-to-fact conditional}. A conditional that presupposes the falsity of its 
antecedent.

{\it Correlata}. The statistically correlated possible results of possible 
measurements performed either on the same system at different times or on 
different systems (see {\it Measurement}).

{\it Counterfactual}. A contrary-to-fact conditional.

{\it Dangle}. Extrinsic properties ``dangle'' from what happens or is the case in 
the rest of the world ${\cal W}-S$ in the sense that they are defined in terms of 
the goings-on in ${\cal W}-S$. They depend for their existence on some actual 
event or state of affairs that can be described without reference to $S$ but has 
implications concerning the possessed properties of $S$. Another way of 
saying this is that the possessed properties of $S$ are supervenient on the 
goings-on in ${\cal W}-S$.

{\it Diachronic correlations}. Correlations between the results of local 
measurements performed on the same system at different times.

{\it Distributed}. Saying that $O$'s position is distributed over a set of mutually 
disjoint regions $\{R_i\vert i=1,\dots,n\}$, is the same as saying that the 
probability of finding $O$ in $R_i$ is positive for several or all values of $i$. If 
the probabilities are subjective, what is distributed (in the literal statistical 
sense) is an ensemble of identically ``prepared'' systems. If the probabilities 
are objective, $O$'s position is not actually but {\it counterfactually} distributed 
(that is, it is distributed over contrary-to-fact conditionals).

{\it EPR correlations}. Correlations between the results of local measurements 
performed on different systems in spacelike separation.

{\it Event}. Something that happens, e.g., the deflection of a pointer needle or 
the click of a detector. Quantum mechanics presupposes actual events or 
states of affairs but does not account for their occurrence or existence. Nor can 
it tell us wherein the actuality of an actual event consists.

{\it Extrinsic property}. A contingent property of $S$ that is undefined, and 
hence not attributable to $S$, unless its being possessed by $S$ can be 
inferred from what happens or is the case in the rest of the world ${\cal W}-S$. 
If $p$ is an extrinsic property, the proposition ${\bf p}=$~``$S$~is~$p$'' is 
meaningless just in case neither its truth nor its falsity can be inferred from the 
goings-on in ${\cal W}-S$.

{\it Factually warranted}. Warranted by a (matter of) fact.

{\it Fuzzy}. Indefinite.

{\it Indefinite}. The value of an observable $A$ (with a specific range of 
possible values $a_i$) is said to be indefinite if there isn't any matter of fact 
from which the actually possessed value can be inferred, and if none of the 
values $a_i$ has an objective probability equal to~1.

{\it Instantaneous state}. The state of a system at an exact time, supposed to be 
evolving in time and to represent not only actually possessed properties but 
also everything of the past that remains causally relevant to the future. 
According to the PIQM, there is no such thing.

{\it Intrinsic property}. A property $p$ of $S$ for which the proposition ${\bf 
p}=$~``$S$~is~$p$'' is ``of itself'' (that is, unconditionally) either true or false at 
any time; neither the truth nor the meaning of $\bf p$ depends on the goings-on 
in the rest of the world ${\cal W}-S$.

{\it Local action ({\rm or} local causality)}. The fallacious notion that local 
matters of fact at a space-time point $\cal P$ depend only on local matters of 
fact in the infinitesimal neighborhood of $\cal P$.

{\it Local realism}. The fallacious notion that physical properties are either 
locally instantiated (that is, they are properties of space-time points) or 
supervenient on locally instantiated properties. It entails the equally fallacious 
notion that space and time (or space-time) are adequately described as point 
sets.

{\it Macroscopic object}. An object whose position is not manifestly indefinite.

{\it Manifestly indefinite}. The distinction between objects whose positions are 
manifestly indefinite, and objects whose positions are not manifestly indefinite, 
is central to the PIQM. It defines a quantum/classical divide without positing ad 
hoc limits to the validity of quantum mechanics, and without compromising on 
the extrinsic nature of all contingent properties. An object $O$ has a manifestly 
indefinite position if the indefiniteness of its position is (counterfactually) 
evidenced by the existence of unpredictable matters of fact about its position. 
The position of a macroscopic object $M$ is not manifestly indefinite: Nothing 
ever indicates that $M$ has a position different from what is predictable on the 
basis of the pertinent classical laws and earlier position-indicating matters of 
fact.

{\it (Matter of) fact}. An actual event or an actual state of affairs.

{\it Measurement}. The existence of a matter of fact (or the occurrence of an 
actual event, or the coming into existence of an actual state of affairs) that 
warrants attributing a specific property to a physical system.

{\it Measurement result}. A matter of fact about the possessed properties of a 
physical system $S$; an actual state of affairs from which the possession by 
$S$ of a specific property can be inferred; a property the possession of which 
is factually warranted.

{\it Nomologically possible}. Consistent with the laws of physics.

{\it Nomological necessity}. The idea that the regularities displayed by empirical 
data are necessitated by causal laws; that things behave the way they do 
because of some invisible causal links. Since all empirical regularities are 
essentially statistical, this idea is an unwarranted anthropocentric projection 
deriving from our self-perception as agents in a successively experienced 
world.

{\it Numerical identity}. Identity proper, to be distinguished from exact similarity.

{\it Objective probability}. A probability assigned to a counterfactual (or to a 
possible result of a not actually performed measurement) on the basis of a 
complete set of relevant facts. Objective probabilities have nothing to do with 
ignorance -- there is nothing to be ignorant of.

{\it Perfect detector}. Between a real detector and a perfect detector there 
exists a qualitative difference that no technological advance can bridge. The 
quantum-mechanical probability associated with finding $O$ in $R_i$, 
$i=1,\dots,n$, is {\it conditional on} the existence of a matter of fact about the 
region containing $O$. Perfect detectors are so defined as to eliminate this 
conditionality. What is a conditional probability for a real detector is an 
absolute probability for a perfect detector.

{\it Positional indefiniteness}. The indefiniteness of a position. The positional 
indefiniteness of a macroscopic object exists only in our minds. Nothing in the 
actual world corresponds to it because it exists only in relation to a background 
space that is more differentiated than the real thing, physical space.

{\it Preparation}. A bad word. It suggests that a measurement prepares an 
evolving actual state of affairs. In reality all that is ``prepared'' by a 
measurement is a set of probability assignments.

{\it Real detector}. The quantum-mechanical probability associated with finding 
$O$ in $R_i$, $i=1,2$, is {\it conditional on} the existence of a matter of fact 
about the region containing $O$. If the quantum-mechanical probability 
associated with finding $O$ in $R_1$ is $1/2$, a perfect detector with sensitive 
region $R_1$ clicks in 50\% of all runs of the experiment, while a real detector 
$D'_1$ with the same sensitive region clicks in 50\% of all runs of the 
experiment {\it in which either detector clicks}. The efficiency of a real detector 
(that is, its likelihood to click when the corresponding Born probability is~1) can 
be measured, but it cannot be calculated from ``first principles,'' for the same 
reason that the occurrence of a causal primary cannot be predicted.

{\it Retrocausation}. Backward-in-time causation.

{\it Retroparation, retropare}. The time reverse of ``preparation'' and ``prepare.''

{\it Separability}. Based on the fallacious notion that space-time is adequately 
described as a point set, the equally fallacious notion that events or states of 
affairs in spacelike separation are independent not only causally but also 
statistically.

{\it Sharp}. Definite. No position is completely sharp. Some positions are 
sharper than others. Some objects (called ``macroscopic'') have the sharpest 
positions in existence; their positional indefiniteness exists only in our minds.

{\it State}. The worst word in the vocabulary of quantum mechanics. 
Legitimately, a state is a collection of actually possessed properties. A quantum 
state, instead, is a collection of (Born) probability assignments. (A ket, like the 
corresponding projection operator, represents a property. If the possession of 
this property is factually warranted at a given time, the ket may be thought of as 
representing this factually warranted property. But the same ket unitarily 
propagated to other times is nothing but a rule for assigning probabilities. To 
keep clear of a potent source of confusion, it is better never to think of a ket as 
representing an actually possessed property.)

{\it State of affairs}. Something that is actually the case, e.g., the needle's 
pointing left. Quantum mechanics presupposes actual events or states of 
affairs but does not account for their occurrence or existence. Nor can it tell us 
wherein the actuality of an actual state of affairs consists.

{\it Subjective probability}. A probability assigned on the basis of an incomplete 
set of relevant facts. If a measurement is actually performed, the probabilities 
associated with the possible results are based on an incomplete set of facts; 
they take no account of the actual result; they therefore express our 
(subjective) ignorance of the actual result.

{\it Supervenient}. See {\it Dangle}.

{\it Temporal referent}. The temporal referent of something $X$ is the time at 
which $X$ happens or obtains. The temporal referent of motion is a set of 
factually warranted moments that is not dense in time.

{\it Trivial probability}. A probability of either 0 or 1.

{\it Truth value}. True or false. A proposition lacks a truth value if it is neither 
true nor false (and therefore meaningless).

{\it Ultimate detector}. A detector whose position and boundary are too sharp to 
be manifestly indefinite.

\bigskip\bigskip\nnref=0\noindent{\bf REFERENCES AND NOTES}

\smallskip\item{a)} E-mail: ujm@auroville.org.in

\def\refitem{\advance\nnref1\item{\number\nnref.}}
\refitem N. David Mermin, ``What is quantum mechanics trying to tell us?,'' Am. 
J. Phys. {\bf 66}, 753-767 (1998).

\refitem David Bohm, {\it Quantum Theory} (Prentice Hall, Englewood 
Cliffs, NJ, 1951).

\refitem I cannot at this point define what I mean by a ``matter of fact about the 
value of an observable,'' except by saying that it is an actual event or state of 
affairs from which that value can be inferred. The question of how to define 
``(matter of) fact,'' ``event,'' ``state of affairs,'' and similar expressions will be 
addressed below.

\refitem Yakir Aharonov, Peter G. Bergmann, and Joel L. Lebowitz, ``Time 
symmetry in the quantum process of measurement,'' Phys. Rev. {\bf 134B}, 
1410-1416 (1964); reprinted in {\it Quantum Theory and Measurement}, edited 
by John Archibald Wheeler and Wojciech Hubert Zurek (Princeton University 
Press, Princeton, NJ, 1983), pp. 680-686.\newcount\wz\wz=\nnref

\refitem Yakir Aharonov and Lev Vaidman, ``Complete description of a quantum 
system at a given time,'' J. Phys. A {\bf 24}, 2315-2328 (1991).

\refitem B. Reznik and Y. Aharonov, `` Time symmetric formulation of quantum 
mechanics,'' Phys. Rev.~A {\bf 52}, 2538-2550 (1995).

\refitem Lev Vaidman, ``Time-symmetrized quantum theory,'' Fortschr. Phys. 
{\bf 46}, 729-739 (1998).

\refitem If the measurement at $t_m$ had yielded $\downarrow_y$, the final 
measurement could not have yielded $\uparrow_y$.

\refitem This point has been forcefully made by Huw Price ({\it Time's Arrow \& 
Archimedes' Point}, Oxford University Press, New York, 1996).

\refitem Professional soccer players and neuroscientists alike may contest this 
account, but that's besides the point.

\refitem Ulrich Mohrhoff, ``Objectivity, retrocausation, and the experiment of 
Englert, Scully and Walther,'' Am. J. Phys. {\bf 67}, 330-335 (1999).

\refitem Michael A.E. Dummett, ``Bringing about the past,'' Philosophical 
Review {\bf 73}, 338-359 (1964).

\refitem Berthold-Georg Englert, Marlan O. Scully and Herbert Walther, ``The 
duality in matter and light,'' Scientific American {\bf 271}, No. 6, 56-61 
(December 1994).

\refitem Marlan O. Scully, Berthold-Georg Englert and Herbert Walther, 
``Quantum optical tests of complementarity,'' Nature {\bf 351}, No. 6322,
111-116 (1991).

\refitem In this context Mermin considers it possible that my now is two weeks 
behind or fifteen minutes ahead of his now. This peculiar notion does not bear 
scrutiny. Temporal relations exist between objective events and/or objective 
states of affairs, not between nows. The use of ``now'' in the plural is at best 
bad English. My experiential now -- the special moment at which the world has 
the technicolor reality it has in my consciousness -- is coextensive with my 
wordline, and so is Mermin's with his worldline. Every event that I have been or 
will be aware of, has had or will have this miraculous kind of reality. Assigning a 
temporal relation to the experiential nows of different persons therefore makes 
as much sense as assigning a temporal relation to two parallel worldines. If I 
point at a spot $(t_1,{\bf x}_1)$ on my worldline and a spot $(t_2,{\bf x}_2)$ on 
your worldline and say, ``When my now is here, yours is there,'' I actually say 
``When my clock shows $t_1$, your clock shows $t_2$.'' But this is a statement 
that makes sense only if it concerns the relation between two coordinate 
systems. As a statement about different times relative to the same coordinate 
system it is a self-contradictory statement about synchronized clocks.

\refitem A. Peres and W.H. Zurek, ``Is quantum theory universally valid?,'' Am. 
J. Phys. {\bf 50}, 807-810 (1982).

\refitem Where real detectors are concerned, we must distinguish between two 
kinds of probability: the probability {\it that} a detector will respond (no matter 
which) and the probability that a specific detector will respond {\it given} that any 
one detector will respond. The latter (conditional) probability is the one that 
quantum mechanics is concerned with. The former (absolute) probability can be 
measured (for instance, by using similar detector in series), but it cannot be 
calculated using the quantum formalism (nor, presumably, any other formalism). 
One can analyze the efficiency of, say, a Geiger counter into the efficiencies of 
its ``component detectors'' (the ionization cross sections of the ionizable targets 
it contains), but the efficiencies of the ``elementary detectors'' cannot be 
analyzed any further. The efficiency of a real detector cannot be calculated from 
``first principles.'' And since the efficiency of a real detector is determined by at 
least one fundamental coupling constant such as the fine structure constant, this 
also implies that a fundamental coupling constant cannot be calculated; it can 
only be gleaned from the experimental data.

\refitem Michael Redhead, {\it Incompleteness, Nonlocality and Realism} 
(Clarendon, Oxford, 1987), p. 72.

\refitem An anonymous referee (of a different paper and a different journal) 
claims that standard quantum mechanics rejects Redhead's sufficiency 
condition but endorses the ``eigenstate-eigenvalue link,'' according to which an 
element of reality corresponding to an eigenvalue of an observable exists at 
time $t$ if and only if the system at $t$ is ``in the corresponding eigenstate of 
this observable.'' It is obvious that the PIQM rejects this claim, since it rejects 
the very notion that quantum states warrant inferences to actualities.

\refitem Asher Peres, ``Can we undo quantum measurements?,'' Phys. Rev. D 
{\bf 22}, 879-883 (1980); reprinted in Wheeler and Zurek (Ref. \number\wz), pp. 
692-696.

\refitem Thus the characterization of a measurement as an ``irreversible act of 
amplification'' is inadequate. As long as what is amplified is counterfactuals, the 
``act of amplification'' is reversible. No amount of amplification succeeds in 
turning a counterfactual into a fact. No matter how many counterfactuals get 
entangled, they remain counterfactuals. On the other hand, once a
property-indicating event or state of affairs has happened or come into 
existence, it is {\it logically} impossible to reverse this. For the relevant fact is 
not that the needle deflects to the left (which could be reversed by returning the 
needle to the neutral position); the relevant fact is that {\it at a time} $t$ the 
needle deflects (or points) to the left. This is a timeless truth. If at the time $t$ 
the needle deflects to the left, then it always has been and always will be true 
that at the time $t$ the needle deflects to the left.

\refitem G. L\"uders, ``\"Uber die Zustands\"anderung durch den 
Messprozess,'' Ann. Physik (Leipzig) {\bf 8}, 322-328 (1951).

\refitem John von Neumann, {\it Mathematical Foundations of Quantum 
Mechanics} (Princeton University Press, Princeton, 1955).

\refitem Wojciech Hubert Zurek, ``Pointer basis of quantum apparatus: Into 
what mixture does the wave packet collapse?'', Phys. Rev.~D {\bf 24},
1516-1525 (1981).

\refitem Wojciech Hubert Zurek, ``Environment-induced superselection rules,'' 
Phys. Rev.~D {\bf 26}, 1862-1880 (1982).

\refitem E. Joos and H.D. Zeh, ``The emergence of classical properties through 
interaction with the environment,'' Zeits. Phys. B -- Condensed Matter {\bf 59}, 
223-243 (1985).

\refitem Wojciech Hubert Zurek, ``Decoherence and the transition from 
quantum to classical,'' Physics Today {\bf 44} (10), 36-44 (1991).

\refitem Wojciech Hubert Zurek, ``Preferred states, predictability, classicality 
and the environment-induced decoherence,'' Prog. Theor. Phys. {\bf 89},
281-312 (1993).

\refitem Robert B. Griffiths, ``Consistent histories and the interpretation of 
quantum mechanics,'' J. Stat. Phys. {\bf 36}, 219-272 (1984).

\refitem M. Gell-Mann and J.B. Hartle, ``Quantum mechanics in the light of 
quantum cosmology,'' in {\it Complexity, Entropy, and the Physics of 
Information}, edited by W.H. Zurek (Addison-Wesley, Reading, MA, 1990), pp. 
425-458.

\refitem Roland Omn\`es, ``Consistent interpretations of quantum mechanics,'' 
Rev. Mod. Phys. {\bf 64}, 339-382 (1992).

\refitem Fay Dowker and Adrian Kent, ``On the consistent histories approach to 
quantum mechanics,'' J. Stat. Phys. {\bf 82}, 1575-1646 (1996).

\refitem N. Gisin and I.C. Percival, ``The quantum-state diffusion model applied 
to open systems,'' J. Phys. A {\bf 25}, 5677-5691 (1992).

\refitem L. Di\'osi, N. Gisin, J.J. Halliwell, and I.C. Percival, ``Decoherent 
histories and quantum state diffusion,'' Phys. Rev. Lett. {\bf 74}, 203-207 
(1994).

\refitem Ian C. Percival, ``Primary state diffusion,'' Proc. R. Soc. Lond. A {\bf 
447}, 189-209 (1994).

\refitem G.C. Ghirardi, A. Rimini, and T. Weber, ``Unified dynamics for 
microscopic and macroscopic systems,'' Physical Review D {\bf 34}, 470-491 
(1986).

\refitem Philip Pearle, ``Combining stochastic dynamical state-vector reduction 
with spontaneous localization,'' Phys. Rev. A {\bf 39}, 2277-2289 (1989).

\refitem Philip Pearle, ``True collapse and false collapse,'' in {\it Quantum 
Classical Correspondence}, edited by Da Hsuan Feng and Bei Lok Hu 
(International Press, Cambridge, MA, 1997), pp. 51-68.

\refitem Jorge Luis Borges, ``The Garden of Forking Paths,'' {\it Ficciones} 
(Everyman's Library, Knopf / Random House, New York, 1993).

\refitem Abner Shimony ``Metaphysical problems in the foundations of quantum 
mechanics,'' International Philosophical Quarterly {\bf 18}, 3-17 (1978).

\refitem Abner Shimony, ``Conceptual Foundations of Quantum Mechanics,'' in 
{\it The New Physics}, edited by Paul Davies (Cambridge University Press, 
Cambridge, 1989), pp. 373-95.

\refitem Werner Heisenberg, {\it Physics and Philosophy} (Harper and Row, 
New York, 1958), Chap.~3.

\refitem Karl R. Popper, {\it Quantum Theory and the Schism in Physics}, edited 
by W.W. Bartley, III (Rowan \& Littlefield, Totowa, NJ, 1982).

\refitem The ``collapse'' of an inference basis is {\it necessarily} unpredictable: 
if it could be predicted, the inference basis would remain unchanged.

\refitem Thomas Nagel, {\it The View from Nowhere} (Oxford University Press, 
New York, 1986).

\refitem Alfred North Whitehead, {\it Process and Reality: An Essay in 
Cosmology} (Macmillan, New York, 1960).

\refitem It is often said that the ``motion'' of the now or the ``flow'' of time are 
purely subjective (Ref.~\number\umretro, Sec.~V). I wouldn't go so far. I prefer 
to think that objective reality encompasses more than ``objective'' science can 
handle. Science knows nothing of the singular and the individual. It deals with 
classes and types and the patterns or regularities that define membership in a 
class. It deals with greylags but not with the greylag goose Martina. It deals 
with lawfulness but not with what instantiates the lawfulness. It deals with the 
laws of physics but not with what it is that obeys the laws of physics. It 
classifies fundamental particles but keeps mum on what a fundamental particle 
intrinsically is. From this it does not follow that, objectively, there is no such 
thing as a fundamental particle. By the same token, from the fact that physics 
can deal only with the quantitative features of time, it does not follow that the 
qualitative features of time are not objective.

\refitem ``...there is no interpolating wave function giving the `state of the 
system' between measurements'' -- Asher Peres, ``What is a state vector?'', 
Am. J. Phys. {\bf 52}, 644-650 (1984).

\refitem Bernard d'Espagnat, {\it Conceptual Foundations of Quantum 
Mechanics}, 2nd ed. (Benjamin, Reading, MA, 1976), p. 251.

\refitem Niels Bohr, {\it Essays 1958-62 on Atomic Physics and Human 
Knowledge} (Wiley, New York, 1963), p. 3.

\refitem Niels Bohr, {\it Atomic Theory and the Description of Nature} 
(Cambridge University Press, Cambridge, 1934).

\refitem Abraham Pais, {\it `Subtle is the Lord...': The Science and the Life of 
Albert Einstein} (Clarendon Press, Oxford, 1982).

\refitem Another way to see this is to recall from note \number\detefcy\ that no 
theoretical account can be given of the efficiency of a real detector -- its 
likelihood to click when the corresponding Born probability is~1. {\it A fortiori}, 
no theoretical account can be given of why or when a detector is certain to 
click. It never is.

\refitem Fritz London and Edmond Bauer, ``The theory of observation in 
quantum mechanics,'' in Wheeler and Zurek (Ref. \number\wz), pp. 217-259.

\refitem Don N. Page, ``Sensible quantum mechanics: Are probabilities only in 
the mind?,'' Int. J. Mod. Phys. D {\bf 5}, 583-596 (1996).

\refitem Henry Pierce Stapp, {\it Mind, Matter, and Quantum Mechanics} 
(Springer, Berlin, 1993).

\refitem Michael Lockwood, {\it Mind, Brain and the Quantum} (Basil Blackwell, 
Oxford, 1989).

\refitem David Z. Albert, {\it Quantum Mechanics and Experience} (Harvard 
University Press, Cambridge, MA, 1992).

\refitem Because these conditions can be stated in classical language, causal 
terms do have a domain of application. More about this in 
Sec.~\secref{\clamac}.

\refitem See Carl Friedrich von Weizs\"acker, {\it The Unity of Nature} (Farrar, 
Straus, Giroux, New York, 1980), Sec.~IV.4.

\refitem B. Misra and E.C.G. Sudarshan, ``The Zeno's paradox in 
quantum theory,'' J. Math. Phys. {\bf 18}, 756-763 (1977).

\refitem C.B. Chiu and E.C.G. Sudarshan, ``Time evolution of unstable 
states and a resolution of Zeno's paradox,'' Phys. Rev. D {\bf 16},
520-529 (1977).

\refitem Asher Peres, ``Zeno paradox in quantum theory,'' Am. J. Phys. 
{\bf 48}, 931-932 (1980).

\refitem N. David Mermin, ``Is the Moon there when nobody looks? Reality and 
the quantum theory,'' Physics Today {\bf 38} (4), 38-47 (1985).

\refitem Departures from the classically predicted positions are necessarily 
random, or unpredictable. A predictable departure would reveal a classical law 
not previously taken into account; it would not be a departure from the 
classically predicted position.

\refitem ``...even when phenomena transcend the scope of classical physical 
theories, the account of the experimental arrangement... must be given in plain 
language, suitably supplemented by technical physical terminology. This is a 
clear logical demand, since the very word `experiment' refers to a situation 
where we can tell others what we have done and what we have learned.'' -- 
Niels Bohr, {\it Atomic Physics and Human Knowledge} (Wiley, New York, 
1958), p. 72.

\refitem If the boundary of a detector $D$ is manifestly fuzzy, there are 
detectors with smaller sensitive regions, so $D$ cannot be among the ultimate 
detectors.

\refitem Albert Einstein, Boris Podolsky, and Nathan Rosen, ``Can 
quantum-mechanical description of physical reality be considered 
complete?,'' Phys. Rev. {\bf 47}, 777-780 (1935); reprinted in Wheeler 
and Zurek (Ref. \number\wz), pp. 138-141.

\refitem Albert Einstein, in {\it Albert Einstein: Philosopher-Scientist}, edited by 
P.A. Schilpp (Open Court, La Salle, Illinois, 1970), p. 85.

\refitem The denial of nomological necessity in physics (sometimes referred to 
as ``causal nihilism'') does not entail that our self-perception as causal agents 
is a delusion. See Ulrich Mohrhoff, ``Interactionism, energy conservation, and 
the violation of physical laws,'' Physics Essays {\bf 10}, 651-665 (1997); ``The 
physics of interactionism,'' Journal of Consciousness Studies {\bf 6}, No. 8/9, 
165-184 (1999).

\refitem ``I think it is safe to say that no one understands quantum mechanics.... 
Do not keep saying to yourself, if you can possibly avoid it, `But how can it be 
like that?' because you will go `down the drain' into a blind alley from which 
nobody has yet escaped. Nobody knows how it can be like that.'' -- Richard P. 
Feynman, {\it The Character of Physical Law} (MIT Press, Cambridge, MA, 
1967), p. 129.

\refitem David K. Lewis, {\it Philosophical Papers, Volume II} (Oxford University 
Press, New York, 1986), p. x.

\refitem Reference~\number\albert, p. 126.

\refitem For a summary see, for instance, Ref.~\number\redh, pp. 49-51; Barry 
Loewer, ``Copenhagen versus Bohmian interpretations of quantum theory,'' 
Brit. J. Phil. Sci. {\bf 49}, 317-328 (1998).

\refitem See, for instance, Rudolf Peierls, ``In defence of 
`measurement','' Physics World {\bf 4} (1), 19-20 (1991).

\refitem Henry Pierce Stapp, ``The Copenhagen interpretation,'' Am. J. Phys. 
{\bf 40}, 1098-1116 (1972).

\refitem N. David Mermin, ``What's wrong with this sustaining myth?'', 
Physics Today {\bf 49} (3), 11-13 (1996).

\refitem Reference~\number\redh, p. 48.
\bye